\documentclass[prl,numbers,a4paper,twoside,secnumarabic,twocolumn,nofootinbib,nobibnotes,showpacs,amsmath,amssymb,floatfix,
italian,frenchb,german,latin,swedish,british
]{revtex4}
\usepackage[T1]{fontenc} 
\usepackage{textcomp}
\renewenvironment{acknowledgements}{\section*{Acknowledgements}}{\par}
\usepackage{babel} 
\newcommand{\langfrench}{\foreignlanguage{french}}
\newcommand{\langgerman}{\foreignlanguage{german}}
\newcommand{\langitalian}{\foreignlanguage{italian}}

\newcommand{\langlatin}{\foreignlanguage{nohyphenation}}
\newcommand{\latin}[1]{{\langlatin{#1}}}

\usepackage{amsthm}
\usepackage[dvipdf]{graphicx}

\usepackage{url}
\usepackage{txfonts}

\DeclareFontFamily{U}{egreek}{\skewchar\font'177}%
\DeclareFontShape{U}{egreek}{m}{n}{<-6>s*[1]eurm5<6-8>s*[1]eurm7 <8->s*[1]eurm10}{}%
\DeclareFontShape{U}{egreek}{m}{it}{<->s*[1]eurmo10}{}%
\DeclareFontShape{U}{egreek}{b}{n}{<-6>s*[1]eurb5 <6-8>s*[1]eurb7 <8->s*[1]eurb10}{}%
\DeclareFontShape{U}{egreek}{b}{it}{<->s*[1]eurbo10}{}

\DeclareFontFamily{U}{egreek}{\skewchar\font'177}%
\DeclareFontShape{U}{egreek}{m}{n}{<-6>s*[1]eurm5<6-8>s*[1]eurm7 <8->s*[1]eurm10}{}%
\DeclareFontShape{U}{egreek}{m}{it}{<->s*[1]eurmo10}{}%
\DeclareFontShape{U}{egreek}{b}{n}{<-6>s*[1]eurb5 <6-8>s*[1]eurb7 <8->s*[1]eurb10}{}%
\DeclareFontShape{U}{egreek}{b}{it}{<->s*[1]eurbo10}{}%
\DeclareSymbolFont{egreeki}{U}{egreek}{m}{it}%
\SetSymbolFont{egreeki}{bold}{U}{egreek}{b}{it}
\DeclareSymbolFontAlphabet{\mathegri}{egreeki}%
\DeclareMathSymbol{\epartial}{\mathalpha}{egreeki}{"40}
\DeclareMathSymbol{\ealpha}{\mathalpha}{egreeki}{"0B}
\DeclareMathSymbol{\ebeta}{\mathalpha}{egreeki}{"0C}
\DeclareMathSymbol{\egamma}{\mathalpha}{egreeki}{"0D}
\DeclareMathSymbol{\edelta}{\mathalpha}{egreeki}{"0E}
\DeclareMathSymbol{\eepsilon}{\mathalpha}{egreeki}{"0F}
\DeclareMathSymbol{\ezeta}{\mathalpha}{egreeki}{"10}
\DeclareMathSymbol{\eeta}{\mathalpha}{egreeki}{"11}
\DeclareMathSymbol{\etheta}{\mathalpha}{egreeki}{"12}
\DeclareMathSymbol{\eiota}{\mathalpha}{egreeki}{"13}
\DeclareMathSymbol{\ekappa}{\mathalpha}{egreeki}{"14}
\DeclareMathSymbol{\elambda}{\mathalpha}{egreeki}{"15}
\DeclareMathSymbol{\emu}{\mathalpha}{egreeki}{"16}
\DeclareMathSymbol{\enu}{\mathalpha}{egreeki}{"17}
\DeclareMathSymbol{\exi}{\mathalpha}{egreeki}{"18}
\DeclareMathSymbol{\eomicron}{\mathalpha}{egreeki}{"6F}
\DeclareMathSymbol{\epi}{\mathalpha}{egreeki}{"19}
\DeclareMathSymbol{\erho}{\mathalpha}{egreeki}{"1A}
\DeclareMathSymbol{\esigma}{\mathalpha}{egreeki}{"1B}
\DeclareMathSymbol{\etau}{\mathalpha}{egreeki}{"1C}
\DeclareMathSymbol{\eupsilon}{\mathalpha}{egreeki}{"1D}
\DeclareMathSymbol{\ephi}{\mathalpha}{egreeki}{"1E}
\DeclareMathSymbol{\echi}{\mathalpha}{egreeki}{"1F}
\DeclareMathSymbol{\epsi}{\mathalpha}{egreeki}{"20}
\DeclareMathSymbol{\eomega}{\mathalpha}{egreeki}{"21}
\DeclareMathSymbol{\evarepsilon}{\mathalpha}{egreeki}{"22}
\DeclareMathSymbol{\evartheta}{\mathalpha}{egreeki}{"23}
\DeclareMathSymbol{\evarpi}{\mathalpha}{egreeki}{"24}

\DeclareMathSymbol{\evarphi}{\mathalpha}{egreeki}{"27}
\DeclareMathSymbol{\evarAlpha}{\mathalpha}{egreeki}{"41}
\DeclareMathSymbol{\evarBeta}{\mathalpha}{egreeki}{"42}
\DeclareMathSymbol{\evarGamma}{\mathalpha}{egreeki}{"00}
\DeclareMathSymbol{\evarDelta}{\mathalpha}{egreeki}{"01}
\DeclareMathSymbol{\evarEpsilon}{\mathalpha}{egreeki}{"45}
\DeclareMathSymbol{\evarZeta}{\mathalpha}{egreeki}{"5A}
\DeclareMathSymbol{\evarEta}{\mathalpha}{egreeki}{"48}
\DeclareMathSymbol{\evarTheta}{\mathalpha}{egreeki}{"02}
\DeclareMathSymbol{\evarIota}{\mathalpha}{egreeki}{"49}
\DeclareMathSymbol{\evarKappa}{\mathalpha}{egreeki}{"4B}
\DeclareMathSymbol{\evarLambda}{\mathalpha}{egreeki}{"03}
\DeclareMathSymbol{\evarMu}{\mathalpha}{egreeki}{"4D}
\DeclareMathSymbol{\evarNu}{\mathalpha}{egreeki}{"4E}
\DeclareMathSymbol{\evarXi}{\mathalpha}{egreeki}{"04}
\DeclareMathSymbol{\evarOmicron}{\mathalpha}{egreeki}{"4F}
\DeclareMathSymbol{\evarPi}{\mathalpha}{egreeki}{"05}
\DeclareMathSymbol{\evarRho}{\mathalpha}{egreeki}{"50}
\DeclareMathSymbol{\evarSigma}{\mathalpha}{egreeki}{"06}
\DeclareMathSymbol{\evarTau}{\mathalpha}{egreeki}{"54}
\DeclareMathSymbol{\evarUpsilon}{\mathalpha}{egreeki}{"07}
\DeclareMathSymbol{\evarPhi}{\mathalpha}{egreeki}{"08}
\DeclareMathSymbol{\evarChi}{\mathalpha}{egreeki}{"58}
\DeclareMathSymbol{\evarPsi}{\mathalpha}{egreeki}{"09}
\DeclareMathSymbol{\evarOmega}{\mathalpha}{egreeki}{"0A} 
\DeclareFontFamily{U}{egreek}{\skewchar\font'177}%
\DeclareFontShape{U}{egreek}{m}{n}{<-6>s*[1]eurm5<6-8>s*[1]eurm7 <8->s*[1]eurm10}{}%
\DeclareFontShape{U}{egreek}{m}{it}{<->s*[1]eurmo10}{}%
\DeclareFontShape{U}{egreek}{b}{n}{<-6>s*[1]eurb5 <6-8>s*[1]eurb7 <8->s*[1]eurb10}{}%
\DeclareFontShape{U}{egreek}{b}{it}{<->s*[1]eurbo10}{}%
\DeclareSymbolFont{egreekr}{U}{egreek}{m}{n}%
\SetSymbolFont{egreekr}{bold}{U}{egreek}{b}{n}
\DeclareSymbolFontAlphabet{\mathegrr}{egreekr}%
\DeclareMathSymbol{\epartialup}{\mathalpha}{egreekr}{"40}

\DeclareMathSymbol{\ealphaup}{\mathalpha}{egreekr}{"0B}
\DeclareMathSymbol{\ebetaup}{\mathalpha}{egreekr}{"0C}
\DeclareMathSymbol{\egammaup}{\mathalpha}{egreekr}{"0D}
\DeclareMathSymbol{\edeltaup}{\mathalpha}{egreekr}{"0E}
\DeclareMathSymbol{\eepsilonup}{\mathalpha}{egreekr}{"0F}
\DeclareMathSymbol{\ezetaup}{\mathalpha}{egreekr}{"10}
\DeclareMathSymbol{\eetaup}{\mathalpha}{egreekr}{"11}
\DeclareMathSymbol{\ethetaup}{\mathalpha}{egreekr}{"12}
\DeclareMathSymbol{\eiotaup}{\mathalpha}{egreekr}{"13}
\DeclareMathSymbol{\ekappaup}{\mathalpha}{egreekr}{"14}
\DeclareMathSymbol{\elambdaup}{\mathalpha}{egreekr}{"15}
\DeclareMathSymbol{\emuup}{\mathalpha}{egreekr}{"16}
\DeclareMathSymbol{\enuup}{\mathalpha}{egreekr}{"17}
\DeclareMathSymbol{\exiup}{\mathalpha}{egreekr}{"18}
\DeclareMathSymbol{\eomicronup}{\mathalpha}{egreekr}{"6F}
\DeclareMathSymbol{\epiup}{\mathalpha}{egreekr}{"19}
\DeclareMathSymbol{\erhoup}{\mathalpha}{egreekr}{"1A}
\DeclareMathSymbol{\esigmaup}{\mathalpha}{egreekr}{"1B}
\DeclareMathSymbol{\etauup}{\mathalpha}{egreekr}{"1C}
\DeclareMathSymbol{\eupsilonup}{\mathalpha}{egreekr}{"1D}
\DeclareMathSymbol{\ephiup}{\mathalpha}{egreekr}{"1E}
\DeclareMathSymbol{\echiup}{\mathalpha}{egreekr}{"1F}
\DeclareMathSymbol{\epsiup}{\mathalpha}{egreekr}{"20}
\DeclareMathSymbol{\eomegaup}{\mathalpha}{egreekr}{"21}
\DeclareMathSymbol{\evarepsilonup}{\mathalpha}{egreekr}{"22}
\DeclareMathSymbol{\evarthetaup}{\mathalpha}{egreekr}{"23}
\DeclareMathSymbol{\evarpiup}{\mathalpha}{egreekr}{"24}

\DeclareMathSymbol{\evarphiup}{\mathalpha}{egreekr}{"27}
\DeclareMathSymbol{\eAlpha}{\mathalpha}{egreekr}{"41}
\DeclareMathSymbol{\eBeta}{\mathalpha}{egreekr}{"42}
\DeclareMathSymbol{\eGamma}{\mathalpha}{egreekr}{"00}
\DeclareMathSymbol{\eDelta}{\mathalpha}{egreekr}{"01}
\DeclareMathSymbol{\eEpsilon}{\mathalpha}{egreekr}{"45}
\DeclareMathSymbol{\eZeta}{\mathalpha}{egreekr}{"5A}
\DeclareMathSymbol{\eEta}{\mathalpha}{egreekr}{"48}
\DeclareMathSymbol{\eTheta}{\mathalpha}{egreekr}{"02}
\DeclareMathSymbol{\eIota}{\mathalpha}{egreekr}{"49}
\DeclareMathSymbol{\eKappa}{\mathalpha}{egreekr}{"4B}
\DeclareMathSymbol{\eLambda}{\mathalpha}{egreekr}{"03}
\DeclareMathSymbol{\eMu}{\mathalpha}{egreekr}{"4D}
\DeclareMathSymbol{\eNu}{\mathalpha}{egreekr}{"4E}
\DeclareMathSymbol{\eXi}{\mathalpha}{egreekr}{"04}
\DeclareMathSymbol{\eOmicron}{\mathalpha}{egreekr}{"4F}
\DeclareMathSymbol{\ePi}{\mathalpha}{egreekr}{"05}
\DeclareMathSymbol{\eRho}{\mathalpha}{egreekr}{"50}
\DeclareMathSymbol{\eSigma}{\mathalpha}{egreekr}{"06}
\DeclareMathSymbol{\eTau}{\mathalpha}{egreekr}{"54}
\DeclareMathSymbol{\eUpsilon}{\mathalpha}{egreekr}{"07}
\DeclareMathSymbol{\ePhi}{\mathalpha}{egreekr}{"08}
\DeclareMathSymbol{\eChi}{\mathalpha}{egreekr}{"58}
\DeclareMathSymbol{\ePsi}{\mathalpha}{egreekr}{"09}
\DeclareMathSymbol{\eOmega}{\mathalpha}{egreekr}{"0A}

\usepackage{bm}
\providecommand{\nequiv}{\not\equiv}
\DeclareMathOperator{\tr}{tr}
\newcommand{\incr}{\Delta}
\newcommand{\defin}{\stackrel{_\text{def}}{=}}

\newcommand{\lmatimplies}{\Rightarrow}

\newcommand{\corr}{\mathrel{\Hat{=}}}
\renewcommand{\le}{\leqslant}
\renewcommand{\ge}{\geqslant}
\DeclareMathDelimiter{\lclose}{\mathopen}{operators}{"5B}{largesymbols}{"02}
\DeclareMathDelimiter{\rclose}{\mathclose}{operators}{"5D}{largesymbols}{"03}
\DeclareMathDelimiter{\lopen}{\mathopen}{operators}{"5D}{largesymbols}{"03}
\DeclareMathDelimiter{\ropen}{\mathclose}{operators}{"5B}{largesymbols}{"02}

\newcommand{\abs}[1]{\lvert#1\rvert}
\newcommand{\ket}[1]{\lvert#1\rangle}

\newcommand{\braket}[2]{\langle#1\mid#2\rangle}
\newcommand{\ketbra}[2]{\lvert#1\rangle\langle#2\rvert}
\newcommand{\proj}[1]{\ketbra{#1}{#1}}

\newcommand{\set}[1]{\{#1\}}

\theoremstyle{definition}

\newcommand{\QED}{\textsc{q.e.d.}}

\newcommand{\chap}{chap.}
\newcommand{\sect}{\S}
\newcommand{\sects}{\S\S}
\newcommand{\eqn}{{Eqn.}}
\newcommand{\eqns}{{Eqns.}}
\newcommand{\etc}{{etc.}}

\newcommand{\ie}{{i.e.}}

\newcommand{\eg}{{e.g.}}
\newcommand{\viz}{{viz.}}
\newcommand{\cf}{{cf.}}
\newcommand{\Cf}{{Cf.}}

\newcommand{\etal}{{et al.}}
\newcommand{\id}{\bm{I}}
\newcommand{\bd}{\hspace{0pt}}%
\newcommand{\povm}{positive-\bd operator-\bd valued measure}
\newcommand{\POVM}{POVM}

\newcommand{\cpm}{completely positive map}
\newcommand{\CPM}{CPM}

\newcommand{\boltzc}{k}
\newcommand{\avogn}{N_{\textrm{A}}}
\newcommand{\zT}{T}
\newcommand{\zN}{N}
\newcommand{\zNp}{n}
\newcommand{\zk}{\boltzc}
\newcommand{\zVf}{V_\text{f}}
\newcommand{\zVi}{V_\text{i}}
\newcommand{\zrho}{\bm{\varrho}}
\newcommand{\zphi}{\bm{\phi}}
\newcommand{\zpsi}{\bm{\psi}}
\newcommand{\zP}{\bm{\varPi}}
\newcommand{\zA}{\bm{A}}
\newcommand{\zE}{\bm{E}}
\newcommand{\zF}{\bm{F}}
\newcommand{\zzp}{z^+}
\newcommand{\zzm}{z^-}
\newcommand{\zzpm}{z^\pm}
\newcommand{\zxp}{x^+}
\newcommand{\zxm}{x^-}
\newcommand{\zxpm}{x^\pm}
\newcommand{\zpm}{\bm{z}^\pm}
\newcommand{\zz}{\bm{z}^+}
\newcommand{\zzb}{\bm{z}^-}
\newcommand{\zx}{\bm{x}^+}
\newcommand{\zzk}{\ket{\zzp}}
\newcommand{\zxk}{\ket{\zxp}}
\newcommand{\zmix}{\bm{\lambda}}
\newcommand{\zab}{\bm{\alpha}^{\pm}}
\newcommand{\za}{\bm{\alpha}^{+}}
\newcommand{\zb}{\bm{\alpha}^{-}}
\newcommand{\zabp}{\alpha^\pm}
\newcommand{\zap}{\alpha^+}
\newcommand{\zbp}{\alpha^-}
\newcommand{\zabk}{\ket{\alpha^\pm}}
\newcommand{\primeds}{\Tilde}
\newcommand{\doubleds}{\Hat}
\newcommand{\zzup}{z^+\, z^+}
\newcommand{\zzupm}{z^\pm\, z^+}
\newcommand{\zzvp}{z^+\, z^-}
\newcommand{\zzvpm}{z^\pm\, z^-}
\newcommand{\zxvp}{x^+\, z^-}
\newcommand{\zxvpm}{x^\pm\, z^-}
\newcommand{\zxup}{x^+\, z^+}
\newcommand{\zxupm}{x^\pm\, z^+}
\newcommand{\zzum}{z^-\, z^+}
\newcommand{\zzvm}{z^-\, z^-}
\newcommand{\zxvm}{x^-\, z^-}
\newcommand{\zxum}{x^-\, z^+}
\newcommand{\zzu}{\primeds{\bm{z}}^{+}}
\newcommand{\zxv}{\doubleds{\bm{x}}^{+}}
\newcommand{\zmi}{\bm{\tau}}
\newcommand{\zwbB}{{{\alpha}}^{\pm}\, z^+}
\newcommand{\zwbBk}{\alpha^\pm\, z^+}
\newcommand{\zwbBpk}{\alpha^+\, z^+}
\newcommand{\zwbBmk}{{\alpha}^-\, z^+}
\newcommand{\zwcC}{{{\alpha}}^{\pm}\, z^-}
\newcommand{\zwcCk}{{\alpha}^\pm\, z^-}
\newcommand{\zwcCpk}{{\alpha}^+\, z^-}
\newcommand{\zwcCmk}{{\alpha}^-\, z^-}
\newcommand{\zwEF}{\bm{E}^{\pm}}
\newcommand{\zaup}{\zwbBpk}
\newcommand{\zavp}{\zwcCpk}
\newcommand{\zbup}{\zwbBmk}
\newcommand{\zbvp}{\zwcCmk}
\newcommand{\xxx}{\bm{\varrho}}

\newcommand{\zAa}{{}^a\text{Ar}}
\newcommand{\zAb}{{}^b\text{Ar}}

\newcommand{\diaphragm}{diaphragm}

\newcommand{\citey}{\citep}
\newcommand{\citeye}{\citep}
\newcommand{\citete}[1]{(#1)}
\newcommand{\citeap}{\citep}
\begin{document}
\bibliographystyle{apsrevmana}

\hyphenation{
im-pli-cans
im-pli-cate
dis-tin-guish-abil-i-ty
prot-a-sis
apod-o-sis
}

\author{\firstname{Piero\,G.\,Luca} \surname{Mana}}
\email{mana@imit.kth.se}

\author{\firstname{Anders} \surname{M\aa{}nsson}}

\author{\firstname{Gunnar} \surname{Bj\"ork}}

\affiliation{School of Information and Communication Technology, 
Royal Institute of Technology (KTH), 
Isafjordsgatan 22, SE-164\,40 Stockholm,
Sweden}


\title{On distinguishability, orthogonality, and violations of
the second law:\\
contradictory assumptions, contrasting pieces of knowledge}


\date{24 August 2005}

\begin{abstract}
Two statements by von~Neumann and a thought-experiment by Peres
prompts a discussion on the notions of one-shot
distinguishability, orthogonality, semi-permeable \diaphragm, and
their thermodynamic implications. In the first part of the paper,
these concepts are defined and discussed, and it is explained
that one-shot distinguishability and orthogonality are
contradictory assumptions, from which one cannot rigorously draw
any conclusion, concerning \eg\ violations of the second law of
thermodynamics. In the second part, we analyse what happens when
these contradictory assumptions comes, instead, from \emph{two}
different observers, having different pieces of knowledge about a
given physical situation, and using incompatible density matrices
to describe it.
\end{abstract}

\maketitle

\section{Introduction: von~Neumann (and Peres) on orthogonality 
and the second law}

In \sect~V.2 of von~Neumann's \emph{\langgerman{{Mathematische}
{Grundlagen} der
{Quantenmechanik}}}~\citep{vonneumann1932c_r1996} (see also
\citeap{vonneumann1927c}) we find the two following
propositions:\footnote{``\langgerman{[Z]wei Zust\"ande $\zphi$,
$\zpsi$ [\ldots]\ durch semipermeable W\"ande bestimmt getrennt
werden k\"onnen, wenn sie orthogonal sind}'', and
``\langgerman{[S]ind $\zphi$, $\zpsi$ nicht orthogonal, so
widerspricht die Annahme einer solchen semipermeablen Wand dem
zweiten Hauptsatz''}.}
\begin{equation}\label{eq:first}
\parbox{.8\columnwidth}{``[T]wo states $\zphi$, $\zpsi$
[\ldots]\ can certainly be separated by a semi-permeable
\diaphragm\ if they are orthogonal'';}
\end{equation}
and the converse
\begin{equation}\label{eq:second}
\parbox{.8\columnwidth}{``[I]f $\zphi$, $\zpsi$ are not
orthogonal, then the assumption of such a semi-permeable
\diaphragm\ contradicts the second law [of
thermodynamics]''.}
\end{equation}

These statements concern thermodynamics and the notions of
\emph{distinguishability} and \emph{orthogonality}. 

Von~Neumann proved the first statement above in an analysis
involving ``thermodynamic considerations'' (the same
considerations by which he derived his entropy formula). However,
he did not actually prove the second statement, but rather the
converse of the first one, \viz: if two ``states'' can be
separated by semi-permeable \diaphragm s, then they must be
orthogonal.

It was Peres~\citey{peres1995} that gave a seemingly direct
demonstration of the second statement~\eqref{eq:second} by means
of a thought-experiment with `quantum gases' in which he
explicitly assumes the non-orthogonality of two quantum
``states'', their separability by semi-permeable \diaphragm s,
and from this assumptions creates a thermodynamic cycle which
violates the second law. More precisely, the thought-experiment
involves \emph{two} observers: one making the non-orthogonality
assumption, the other performing the separation.

The interplay of orthogonality, distinguishability,
thermodynamics, and multiplicity of observers is quite
interesting; therefore we want to discuss and analyse it in
varying depth, mainly with paedagogical purposes. The paper is
divided into two main parts, reflecting two main perspectives.

In the first part we offer a discussion on the three concepts of
\emph{preparation} and \emph{measurement procedure} and
\emph{``one-shot'' distinguishability}, and of their mathematical
representatives: \emph{density matrix}, \emph{\povm}, and
\emph{orthogonality}. First, we make clear that `orthogonality'
is simply the mathematical counterpart of `one-shot
distinguishability'; thus, one cannot assert or deny the one
without asserting or denying the other as well. Second, we show
how the idea of a semi-permeable membrane which can separate two
preparations (``states'') is just a particular realisation of a
measurement procedure which can distinguish, in one shot, those
preparations. This will provide an occasion to discuss the
relation of these concepts to thermodynamics. The principal
conclusion of the first part will be that the second
statement~\eqref{eq:second} --- and its demonstration by Peres
--- contains a contradiction in its premise; \ie, it has the
logical form `$(A \land \lnot A) \lmatimplies B$'. The
contradiction in the premise is the assumption of
distinguishability and non-distinguishability at the same time.

The contradiction in Peres' experiment, however, can also be
conceived as a situation in which \emph{two} scientists use two
\emph{different} density matrices to analyse the \emph{same}
physical phenomenon. This provides an illustration of the fact
that a density matrix adopted to describe a preparation is always
dependent on an observer's particular \emph{knowledge} about that
preparation~\citep{kemble1939,kemble1939b,jaynes1979b,jaynes1993b,jaynes1989,jaynes1990b,jaynes2003,fuchs2001,fuchsetal2004,brunetal2002,cavesetal2002b,cavesetal2002c,mana2003,mana2004b,mana2005},
as well as on the sets of preparations and measurements
considered by the
observer~\citep{peres1995,hardy2001,mana2003,mana2004b,mana2005}.
It also illustrates what can happen when the two observers' have
and \emph{use} different pieces of knowledge, hence different
density-matrix assignments. For example, the second law may
seemingly be violated for one of the observers. The fact that a
mathematical description always depends on one's particular
knowledge of a phenomenon, however, is not only true in quantum
mechanics, but in classical physics as well. A particular case
within thermodynamics was shown by Jaynes'~\citey{jaynes1992} by
means of a thought-experiment which is very similar to Peres' and
in which the same seeming violation of the second law appears
according to one of the observers. For this reason, we shall
juxtapose Peres' demonstration to Jaynes', hoping that they will
provide insight into each other.


We assume that our reader has a working knowledge of quantum
mechanics and thermodynamics (we also provide some footnotes on
recent developments of the latter, today better called
`thermomechanics'~\citep{truesdell1969_r1984,samohyl1987,silhavy1997},
since they are apparently largely unknown to the quantum-physics
community.)  We also want to emphasise that this paper is not
directly concerned with questions about the relation between
thermodynamics and statistical mechanics, nor to questions about
``classical'' or ``quantum'' entropy formulae. Since our
discussions will regard the second law of thermodynamics, the
reader probably expects that entropy will enter the scene; but
his or her expectations will not be fulfilled. Peres',
von~Neumann's, and Jaynes' demonstrations are based on cyclic
processes, which start from and end in a situation described by
the same \emph{thermodynamic state}.\footnote{The concept of
`state' in thermomechanics may include not only the instantaneous
values of several (field) variables, but even their histories, or
suitably defined equivalence classes thereof. This concept has an
interesting historical development. \Cf\ \eg\ Truesdell
\etal~\citep[Preface]{truesdelletal1965_r1992},
Noll~\citep{noll1972,noll2004}, Willems~\citep{willems1972},
Coleman
\etal~\citep{colemanetal1974,colemanetal1975,colemanetal1977,colemanetal1979}
Del~Piero \etal~\citep{delpiero1985,delpieroetal1997},
\v{S}ilhav\'y~\citep{silhavy1997}.} Assuming the entropy to be a
state variable, its change is naught in such processes,
independently of its mathematical expression. The second law of
thermodynamics assumes then the form\footnote{See
Serrin~\citey{serrin1979} for a keen analysis of the second law
for cyclic processes.}
\begin{equation}
\label{eq:seclawcycl}
Q \le 0 \quad
\text{(cyclic processes)},
\end{equation}
where $Q$ is the total amount of heat \emph{absorbed} by the
thermodynamic body in the process. This is the entropy-free form
we are going to use in this paper.


\section{Part I: Contradictory
premises and violations of the second law}
\label{part:I}
\begin{flushright}
{\footnotesize\begin{minipage}{.7\columnwidth} \emph{Alice felt
even more indignant at this suggestion. ``I mean,'' she said,
``that one can't help growing older.''}~\citep{carroll1871}
\end{minipage}\par}
\end{flushright}

\subsection{Preparations and density matrices, measurements and
\povm s\label{sec:introobjects}}

We begin by informally recalling the definitions of `preparation
procedure' and `density matrix', `measurement procedure' and
`\povm', and `one-shot distinguishability' and
`orthogonality'. In our definitions we follow
Ekstein~\citey{ekstein1967,ekstein1969},
Giles~\citey{giles1968,giles1970,giles1979}, Foulis and
Randall~\citep{foulisetal1972a,foulisetal1978,randalletal1973b},
Band and Park~\citep{parketal1976,bandetal1976}, and
Peres~\citey{peres1984,peres1986,peres1995}; the reader is
referred to these references for a deeper discussion \citete{see 
also~\citeap{kemble1939,kemble1939b,ballentine1970,stapp1971,stapp1972,peres1978,kraus1983,hardy2001,mana2004b,mana2003}
and the introductory remarks in Komar~\citeye{komar1962}}.


In quantum theory, a \emph{preparation procedure} (or
`preparation' for short) is ``an experimental procedure that is
completely specified, like a recipe in a good cookbook''
\citete{\citeap{peres1995}, p. 12, \cf\ also p.~424}. It is
usually accompanied by a \emph{measurement procedure},
which can result in different outcomes, appropriately labeled.
The probabilities that we assign to the obtainment of these
outcomes, for all measurement and preparation procedures, are set
as postulates in the theory and encoded in its mathematical
objects, described below. Preparation and measurement procedures
are not mathematical objects.
As an oversimplified example, the instructions for the set-up and
triggering of a low-intensity laser constitute a preparation
procedure, and the instructions for the installation of a
detector constitute a measurement procedure.

A preparation procedure is \emph{mathematically represented}
by a \emph{density matrix}\footnote{The original and more apt
term was `statistical matrix', which has unfortunately been
changed to `density matrix' apparently since
Wigner~\citey{wigner1932}; \cf\ Fano~\citey{fano1957}.} $\zrho$,
whose mathematical properties we assume well known to the
reader~\citep{peres1995,fano1957,kraus1983}. A measurement
procedure is instead mathematically represented by a \emph{\povm}
(\POVM) $\set{\zE_i}$, \viz, a set of positive (semi-definite)
matrices, each associated to a particular measurement outcome,
which sum to the identity
matrix~\citep{kraus1971,kraus1983,davies1983,peres1995,peres2000a}.
These two mathematical objects encode our knowledge of the
statistical properties of their respective preparation and
measurement: when we perform an instance of the measurement
represented by $\set{\zE_i}$ on an instance of the preparation
represented by $\zrho$, the probability $p_i$ assigned to the
obtainment of outcome $i$ is given by the trace formula $p_i =
\tr(\zrho\zE_i)$.

If we wish to specify not only the probabilities of the various
outcomes of a measurement, but also its effect on the
preparation, we associate to the measurement procedure a
\emph{\cpm}
(\CPM)~\citep{kraus1971,kraus1983,choi1975,davies1983,peres2000a}.
We shall not use the full formalism of \cpm s here, however, but
only the special case where the \CPM\ is a set of projectors
$\set{\zP_i}$ and the effect on the density matrix $\zrho$ when
result $i$ is obtained is given by
\begin{equation}
\label{eq:projeffect}
\zrho \mapsto \zrho_i' = 
\frac{\zP_i \zrho \zP_i}{\tr(\zP_i \zrho \zP_i)}.
\end{equation}

\subsection{One-shot distinguishability and orthogonality}
\label{sec:distorth}


Suppose a physicist has realised \emph{one instance} of a
preparation procedure,
choosing between two possible preparation
procedures.\footnote{Note the difference between a preparation
(or measurement) procedure, and the realisation of an instance
thereof. The first is a set of instructions, a description, the
latter is a single actual realisation of those instructions. This
terminology may help avoiding the confusion which some authors
\citete{see \eg\ 
\citeap{ghirardietal1975,cohen1999,longetal2004}} still make
between `ensemble' and `assembly', as defined and conceptually
distinguished by \eg\ Peres \citep[pp.~25, 59, 292]{peres1995}
(see also Ballentine~\citep[p.~361]{ballentine1970}). The term
`ensemble' is used with so different meanings in the literature
(see \eg\ Hughston \etal~\citey{hughstonetal1993} for yet
another, though self-consistent, usage), that we prefer to avoid
it altogether. Moreover, Bayesian probability theory \citete{see
\eg\ \citeap{hobson1971,jaynes2003}} makes the conceptual
usefulness of this term questionable or at least obsolete.} We do
not know which of the two, but we can perform a measurement on
the instance and record the outcome. Suppose there exists a
measurement procedure such that some (at least one) of its
outcomes have vanishing probabilities for the first preparation
and non-vanishing probabilities for the second, while the
remaining (at least one) outcomes have vanishing probabilities
for the second preparation and non-vanishing probabilities for
the first. This means that by performing \emph{a single instance}
of this measurement procedure we can deduce \emph{with certainty}
which preparation was made, by looking at the outcome. In this
case the two preparations are said to be \emph{one-shot
distinguishable} \citete{\cf\ \citeap{hardy2001}}. The fact that
some sets of preparation procedures are one-shot distinguishable
while others are not is the basic reason why quantum mechanics is
a probabilistic theory.

If two preparation procedures are one-shot distinguishable, it is
then easy to separate them. By this we mean that if we are
presented with many instances of these preparations, we can for
each instance make a measurement and tell the preparation, and
thus separate the instances of the first preparation from the
instances of the second preparation. But the converse is also
true: if we can separate with certainty the two groups of
instances, it means that we can distinguish with certainty the
preparation of each instance. 

Let us see how the one-shot-distinguishability property is
represented mathematically. The two one-shot-distinguishable
preparation procedures of the above example are represented by
density matrices $\zphi$ and $\zpsi$, and the measurement by a
\POVM\ $\set{\zA_k}$. According to the definition of one-shot
distinguishability given above, we must be able to write this
\POVM\ as $\set{\zA_k}= \set{\zE_i, \zF_j}$, with
\begin{equation}
\begin{split}
&\tr(\zphi\zE_i) = 0\quad \text{and} \quad\tr(\zpsi\zE_i) \neq 0
\qquad\text{for all $\zE_i$},\\
&\tr(\zpsi\zF_j) = 0\quad \text{and} \quad\tr(\zphi\zF_j) \neq 0
\qquad\text{for all $\zF_j$}.
\end{split}\label{eq:disting}
\end{equation}
This is the mathematical form of our definition of one-shot
distinguishability.

An important consequence of the equations above is the following:
the density matrices $\zphi$ and $\zpsi$ must be orthogonal,
\viz\ $\tr(\zphi \zpsi) = 0$.\footnote{This expression is the
usual definition of orthogonality, \ie\ vanishing scalar product,
between vectors: in our case the density matrices are considered
as vectors in a real vector space of Hermitean matrices, with
$\tr(\zphi \zpsi)$ as the scalar product.} We prove this simple
theorem in the Appendix. 

Thus, \emph{the fact that two preparation procedures are one-shot
distinguishable 
is mathematically represented by the orthogonality of their
associated density matrices}. Or, to put it another way,
orthogonality is the mathematical representation of one-shot
distinguishability. The converse of the mathematical theorem is
also true: if two density matrices are orthogonal, then there
exists a \POVM\ with the properties~\eqref{eq:disting} above. But
to conclude that the represented preparation procedures are
one-shot distinguishable, we need first to \emph{assume} that
there \emph{physically exists} a measurement procedure
corresponding to that \POVM.\footnote{This is not a light
assumption; \cf\ \eg\ Peres~\citep{peres1995}, pp.~50
and~424.\label{fn:existassump}}.


To prove this theorem (and its converse), thermodynamic arguments
are \emph{not} needed. It just follows mathematically from
\eqns~\eqref{eq:disting}, which represent a probabilistic
property of a measurement. So we begin to see that von~Neumann's
statement~\eqref{eq:first}, in which he seems to derive
orthogonality from separability by semi-permeable \diaphragm s,
would not really need thermodynamic considerations. But we can
make this more precise only after we have discussed the notion of
a \emph{semi-permeable \diaphragm}, which appears in
statement~\eqref{eq:first}. This will be done in a moment; before
then, we want to make some final remarks about distinguishability
and about linguistic details.

It may be the case that two preparation procedures cannot be
distinguished by means of a single measurement instance, but can
still be distinguished with arbitrary precision by studying their
statistical properties, \ie\ by analysing the results of an
adequately large number of diverse measurements on \emph{separate
instances} of these preparations.\footnote{See \eg\ 
\citey{wootters1981,woottersetal1989,peresetal1991,jones1991b,jones1994,braunsteinetal1994,slater1995b,gilletal2000};
\cf\ also \citeap{brunetal2002,cavesetal2002c}.} They are thus
distinguishable, \emph{but not in one-shot}. They are represented
in this case by \emph{different}, but non-orthogonal, density
matrices. On the other hand, two preparation procedures are
represented by the \emph{same} density matrix when they have
exactly the same statistical properties (with respect to a given
set of measurement procedures) and so cannot be distinguished by
a statistical analysis of measurement results, no matter how many
instances of them we prepare and how many measurements we
make.

The whole discussion above is easily generalised to more than two
preparation procedures.

\subsection{Interlude: linguistic details}
\label{sec:lingdet}

Preparation procedures are often called `states'; however, the
word `state' is equally often used to mean the density matrix
representing a preparation procedure. This double meaning is
quite natural, because the concepts of preparation procedure and
density matrix are, as we have seen, strictly related. They are
nevertheless distinct, as shown \eg\ by the fact that
\emph{different} preparation procedures (concerning the same
physical phenomenon) can sometimes be represented by the
\emph{same} density matrix. We wish to keep this distinction in
this paper, and we shall thus stick to the two distinct terms,
avoiding the term `state'.

The terms `(one-shot) distinguishable' 
and `orthogonal' are also often interchanged in an analogous way;
\eg, one says that ``two preparations are orthogonal'' (instead
of ``two preparations are distinguishable''), or that ``two
density matrices are distinguishable'' (instead of ``two density
matrices are orthogonal''). Such metonymic expressions are of
course handy and acceptable,
but we must not forget that `orthogonal preparations' only means
`one-shot distinguishable preparations', so that if we say
``these preparations are one-shot distinguishable and
non-orthogonal'' we are then contradicting ourselves --- not an
experimental or physical contradiction, but a \emph{linguistic},
or \emph{logical}, one. Just as it would be contradictory to say
that a classical force field is conservative and its integral
along a closed path does not vanish; or that, in a given
reference frame, a point-like body is in motion and its
position-vector is constant; or that a non-relativistic system is
closed and its total mass is changing.

These remarks are pedantic, and many a reader will consider them
only linguistic nit-picking; but these readers are then invited
to take again a look at von~Neumann's second statement
\eqref{eq:second}. Is everything in order there? We shall come
back to this point later.

\subsection{``Quantum'' ideal gases and 
semi-permeable \diaphragm s\label{sec:matintro}}

In order to analyse a statement which involves, besides quantum
concepts, also thermodynamic arguments and semi-permeable
\diaphragm s, it is necessary to introduce a thermodynamic body
possessing ``quantum'' characteristics, \ie, quantum degrees of
freedom; an `ideal' body will do very well. For this purpose, we
shall first recall the basic relationships between work and heat
for classical ideal gases. We shall then follow
von~Neumann~\citey{vonneumann1932c_r1996} and introduce the
quantum degrees of freedom as ``internal'' degrees of freedom of
the particles constituting the gases, and shall finally discuss
how the interaction between the quantum and thermodynamic
parameters is achieved by means of semi-permeable \diaphragm s.

Let us first recall that (classical) ideal gases are defined as
homogeneous, uniform thermodynamic bodies characterisable by two
thermodynamic variables: the volume $V>0$ and the temperature
$\zT>0$, and for which the internal energy is a function of the
variable $\zT$ alone;\footnote{See \eg\ the excellent little book
by Truesdell and Bharatha~\citey{truesdelletal1977}, or
Planck~\citep[\sects~86--91, 232--236]{planck1897_t1945}, Lewis
and Randall~\citep[\chap~VI]{lewisetal1923},
Partington~\citep[\sects~II.54--57, IV.14,
VIIA.21]{partington1949}, Callen~\citep[\sects~3-4, 13-1--2,
16-10]{callen1960_r1985}, Buchdahl~\citep[\sects~71,
82]{buchdahl1966}; also Samoh\'yl~\citep[\sects~5,
6]{samohyl1987}.} this implies, via the first law of
thermodynamics, that in any isothermal process the heat
\emph{absorbed by the gas}, $Q$, is always equal to the work
\emph{done by the gas}, $W$:
\begin{gather}
\label{eq:working}
Q=W=\zN \zk \zT \ln (\zVf/\zVi) \quad \text{(isothermal
processes)},
\end{gather}
where $\zVi$ and $\zVf$ are the volumes at the beginning and end
of the process,
$\zk$ is Boltzmann's constant, and $\zN$ is the (constant) number
of particles. This formula will be true throughout the paper, as
we shall only consider isothermal processes. (Note that $W$ and
$Q$ can assume both positive and negative values.)

One often considers several samples of such ideal gases in a
container and inserts, moves, or removes impermeable or
semi-permeable \diaphragm
s\footnote{Partington~\citep[\sect~28]{partington1949} informs us
that these were first used in thermodynamics by
Gibbs~\citey{gibbs1876}.} at any position one
pleases.\footnote{In the limit, this leads
to a formalism involving field quantities \citete{\cf\ 
Buchdahl~\citep[\sects~46, 75]{buchdahl1966}, and see \eg\ 
Truesdell~\citep[lectures~5,~6 and related
appendices]{truesdell1969_r1984}
and~\citeap{truesdelletal1960,bowen1968,truesdell1968b,samohyl1987,samohyl1999}},
as indicated by the possibility of introducing as many \diaphragm
s as we wish and hence to control smaller and smaller portions of
the gas samples.} What the presence of these \diaphragm s, both
in practice and in theory, really signifies in the case of ideal,
non-interacting (\eg\ non chemically reacting) gas samples, is
that we can control and monitor the variables of the ideal-gas
samples, $(V_1, \zT_1; V_2, \zT_2; \dotsc)$ independently of each
other, even when some of the samples occupy identical regions of
space simultaneously. Thus, the problem becomes equivalent to one
where all gas samples always occupy distinct regions of space,
even though they may be in mechanical or thermal
contact.\footnote{In particular, the samples have separate
entropies and must separately satisfy the second
law~\citep{gibbs1876}. When the gases \emph{do} interact, we
enter into the more complex, and still under development,
thermomechanic theory of mixtures~(see \eg\ 
\citep{mueller1968,truesdell1968b,truesdell1969_r1984,bowenetal1970,atkinetal1976,atkinetal1976b,fosdicketal1986,samohyl1987,samohyl1999};
\cf\ also~\citep{ehrenfestetal1920}).} For ``quantum'' ideal
gases, as will be seen in a moment, the situation is not so
simple.

We must first face the question of how to introduce and
mathematically represent quantum degrees of freedom in an ideal
gas. Von~Neumann~\citep[\sect~V.2]{vonneumann1932c_r1996}
used a hybrid classical-quantum description, microscopically
modelling a quantum ideal gas as a large number $\zN$ of
classical particles possessing an ``internal'' quantum degree of
freedom represented by a density matrix $\zrho$ living in an
appropriate density-matrix space. This space and the density
matrix are always assumed to be the same for all the gas
particles.\footnote{Dieks and van~Dijk~\citey{dieksetal1988}
point rightly out that one should then more correctly consider
the total density matrix $\bigotimes_{i=1}^{\zNp}\zrho$, where
$\zNp = \zN \avogn$ (with $\avogn$ Avogadro's constant) is their
total number.\label{fn:dieks}} He then treated two gas samples
described by different density matrices as gases of \emph{somehow
different chemical species}.
This idea had been presented by Einstein~\citey{einstein1914}
eighteen years earlier, but it is important to point out that the
``internal quantum degree of freedom'' was for Einstein just a
``resonator'' capable of assuming only discrete energies. This
degree of freedom had for him discrete (the original meaning of
the adjective `quantum') but otherwise \emph{statistically
classical} properties; it was not described by density matrices,
and it did not provide non-\bd orthogonality issues. Einstein's
idea was hence less open to problems than von~Neumann's.%

In fact, von~Neumann's conceptual device presents some
problems.
For example, the chemical species of a gas is not a thermodynamic
variable, and even less a continuous one: chemical differences
cannot change \emph{continuously} to zero.\footnote{An
observation made by
Partington~\citep[\sect~II.28]{partington1949} in a reference to
Larmor~\citep[p.~275]{larmor1897}.} It would then seem more
appropriate to describe a quantum ideal gas by the variables $(V,
\zT, \zrho)$ instead, taking values on appropriate sets. That
this would indeed be the only correct treatment can be seen from
the fact that the thought experiments considered by von~Neumann
and reproduced below always involve some step in which the
density matrix $\zrho$ of a gas is \emph{changed}. This implies
that $\zrho$ is something which we can and \emph{need to}
control, and as such it should be included in the list of
variables which \emph{define} our thermodynamic
system~\citep{bjoerketal2004b}.

However, in the following discussion we shall follow von~Neumann
and Peres instead and speak of a `$\zphi$-gas', or a
`$\zpsi$-gas', \etc, where $\zphi$ or $\zpsi$ are the density
matrices describing the internal quantum degrees of freedom of
the gas particles, just as if we were speaking of gases of
different chemical species (like \eg\ `argon' and `helium'). The
thermodynamic variables are $(V, \zT)$ for each such gas.

In this framework, semi-permeable \diaphragm s have different
meaning and function than they have in the classical framework.
This becomes clear when we analyse how they are modelled
microscopically. The microscopic picture
\citep[p.~196]{vonneumann1932c_r1996}\citep[p.~271]{peres1995}
is, paraphrasing von~Neumann, to construct many ``windows'' in
the \diaphragm, each of which is made as follows. Each particle
of the gases is detained there and a \emph{measurement} is
performed on its quantum degrees of freedom; depending on the
measurement result, the particle penetrates the window or is
reflected, and its density matrix is \emph{changed} in an
appropriate way. In other words, the \diaphragm\ also acts on the
translational degrees of freedom, \emph{separating} the particles
having different preparations spatially, with an efficiency which
depends on the preparations and the implemented measurement. This
implies that the number of particles and hence the pressures or
volumes of the gases on the two sides of the semi-permeable
\diaphragm\ will vary, and may set the \diaphragm\ in motion,
producing work (\eg\ by lifting a weight which loads the
\diaphragm). There arises thus a kind of mutual dependence
between the quantum degrees of freedom and the thermodynamic
parameters (like the volume $V$) of the quantum ideal gases. We
see that for quantum gases the semi-permeable \diaphragm s not
only make the existence of separate thermodynamic variables for
different gas samples possible, as it happened for
(non-interacting) classical gases, but also perform
transformations of the (not thermodynamically reckoned) quantum
variable $\zrho$.

Such a \diaphragm\ is then simply \emph{a device which implement
a measurement procedure}, and is mathematically described by a
given \POVM\ and a \CPM.

Let us illustrate how semi-permeable \diaphragm s work with two
examples.

\subsubsection{First example: one-shot distinguishable
preparations}
\label{sec:example1}

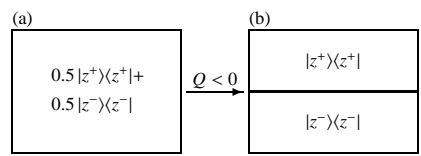
\begin{figure}[b]
\setlength{\unitlength}{0.0047\columnwidth}
\begin{picture}(140,50)(0,0)
\put(0,0){\framebox(55,40){\scriptsize$\begin{aligned}&0.5\,\ketbra{\zzp}{\zzp}+\\&0.5\,\ketbra{\zzm}{\zzm}\end{aligned}$}}
\put(0,41){\makebox(0,0)[bl]{\scriptsize(a)}}

\put(57,20){\vector(1,0){18.5}}
\put(66.25,21){\makebox(0,0)[b]{\scriptsize$Q<0$}}

\put(77.5,20){\framebox(55,20){\scriptsize$\ketbra{\zzp}{\zzp}$}}
\put(77.5,0){\framebox(55,20){\scriptsize$\ketbra{\zzm}{\zzm}$}}
\put(77.5,41){\makebox(0,0)[bl]{\scriptsize(b)}}
\end{picture}
\caption{Separation of one-shot-distinguishable quantum gases.}\label{fig:distin}
\end{figure}

Imagine a container having volume $V$ and containing a mixture of
$\zN/2$ particles of a $\zz$-gas and $\zN/2$ of a $\zzb$-gas;
\ie, the particles have quantum degrees of freedom represented by
the density matrices
\begin{align}
\zz &\defin \ketbra{\zzp}{\zzp} \corr
\Bigl(\begin{smallmatrix} 1&0\\0&0 
\end{smallmatrix}\Bigr),\label{eq:zplus}\\
\zzb &\defin \ketbra{\zzm}{\zzm} \corr
\Bigl(\begin{smallmatrix} 0&0\\0&1
\end{smallmatrix}\Bigr),
\end{align}
in the usual spin-1/2 notation (Fig.~\ref{fig:distin},~a). These
two density matrices represent quantum preparations that can be
distinguished in one shot by an appropriate measurement,
implemented by two semi-permeable \diaphragm s as described
above. The first is completely opaque to the particles of the
$\zzb$-gas and completely transparent to those of the $\zz$-gas;
the other is completely opaque to the particles of the $\zz$-gas
and completely transparent to those of the $\zzb$-gas.
Mathematically they are represented by the two-element \POVM\ 
$\set{\zpm}$ and the \CPM s (projections) $\set{\xxx \mapsto
\zpm\xxx\zpm/ \tr(\zpm\xxx\zpm)}$, as follows.\footnote{Due to
the large number of gas particles considered, the outcome
probabilities are numerically equal, within small fluctuations
negligible in the present work, to the average fraction of gas
correspondingly transmitted or reflected by the \diaphragm s.}
For the first \diaphragm:
\begin{subequations}
\label{eq:actplus}
\begin{align}
\zz &\mapsto 
\zz, \text{ let through, with probability $\tr(\zz\zz)=1$},\\
\zzb &\mapsto 
\zzb, \text{ reflected, with probability
$\tr(\zzb\zzb)=1$}\\
\zz &\mapsto 
\zzb, \text{ with probability $\tr(\zzb\zz)=0$},\\
\zzb &\mapsto 
\zz, \text{ with probability $\tr(\zz\zzb)=0$},
\end{align}
\end{subequations}
and for the second:
\begin{subequations}
\label{eq:actminus}
\begin{align}
\zz &\mapsto 
\zz, \text{ reflected, with probability $\tr(\zz\zz)=1$},\\
\zzb &\mapsto 
\zzb, \text{ let through, with probability
$\tr(\zzb\zzb)=1$}\\
\zz &\mapsto 
\zzb, \text{ with probability $\tr(\zzb\zz)=0$},\\
\zzb &\mapsto 
\zz, \text{ with probability $\tr(\zz\zzb)=0$}.
\end{align}
\end{subequations}

Now imagine we insert these two \diaphragm s in the container
(Fig.~\ref{fig:distin},~a), very near to its top and bottom walls
respectively. We push them isothermally toward the middle of the
container until they come in contact with each other, so that the
container is divided in two chambers. By doing so we have
separated the two gases, with the $\zz$-gas in the upper chamber
and the $\zzb$-gas in the lower one (Fig.~\ref{fig:distin},~b).
In order to move these \diaphragm s and achieve this separation
we have spent an amount of work equal to
\begin{equation}\label{eq:worksepar}
- 2\times \frac{\zN}{2} \zk \zT \ln\frac{V/2}{V}
\approx 0.693\, \zN \zk \zT,
\end{equation}
because each \diaphragm\ had to overcome the pressure exerted by
the gas to which it is opaque. Since the process is isothermal,
the quantity above is also the (positive) amount of heat released
by the gases.

The semi-permeable \diaphragm s can also be used to realise the
inverse process, \ie\ the mixing of two initially separated
$\zz$- and $\zzb$-gases. In this case the
expression~\eqref{eq:worksepar} would be the amount of heat
absorbed by the gases as well as the amount of work performed by
them.

\subsubsection{Second example: non-one-shot distinguishable
preparations}
\label{sec:example2}

The quantum degrees of freedom of two gases may also be prepared
in such a way that no measurement procedure can distinguish
between them in one shot, and so there are no semi-permeable
\diaphragm s which can separate them completely. This has of
course consequences for the amount of work that can be gained by
using the \diaphragm s. Imagine again the initial situation
above, but this time with a mixture of $\zN/2$ particles of a
$\zz$-gas and $\zN/2$ of an $\zx$-gas, with
\begin{equation}
\zx \defin \ketbra{\zxp}{\zxp} \corr \tfrac{1}{2}
\Bigl(\begin{smallmatrix} 1&1\\1&1
\end{smallmatrix}\Bigr).\label{eq:xplus}
\end{equation}
The two density matrices $\zz$ and $\zx$ are non-\bd
orthogonal, $\tr(\zz\zx)\neq 0$, and this must represent the fact
that the preparation procedures they represent cannot be
distinguished in one shot.  This also means that there are no
semi-permeable \diaphragm s which are completely opaque to the
particles of the one gas and completely transparent to those of
the other gas and \latin{vice versa}. Mathematically this is
reflected in the non-\bd existence of a \POVM\ with the
properties~\eqref{eq:disting}.
The absence of such completely separating \diaphragm s implies
that we cannot control their mixing in a reversible way.

It can be shown~\citep{vonneumann1932c_r1996,peres1995} that in
this case the separating process requiring the minimum amount of
work, and so the maximum (negative) amount of heat absorbed by
the gas,\footnote{Readers interested in entropy questions should
note that such process is irreversible and that \emph{there is no
reversible process} with the same initial and final thermodynamic
states, as discussed by Dieks and
van~Dijk~\citey{dieksetal1988}.} can be performed by two
\diaphragm s represented by the two-element \POVM\ $\set{\zab}$
and the \CPM s (projections) $\xxx \mapsto \zab\xxx\zab/
\tr(\zab\xxx\zab)$, with
\begin{align} 
\zab &\defin
\ketbra{\zabp}{\zabp} \corr \frac{1}{4}\begin{pmatrix}
2\pm\sqrt{2}&\pm\sqrt{2}\\ \pm\sqrt{2}& 2\mp \sqrt{2}
\end{pmatrix},\\
\begin{split}\label{eq:newketalf}
\zabk &\defin \Bigl(2\pm \sqrt{2}\Bigr)^{-\frac{1}{2}}
\bigl(\ket{\zzpm} \pm \ket{\zxpm}\bigr) \equiv{} \\
&\qquad
\frac{1}{2}\Bigl[\pm\Bigl(2\pm\sqrt{2}\Bigr)^{\frac{1}{2}}
\ket{\zzp} + \Bigl(2\mp\sqrt{2}\Bigr)^{\frac{1}{2}}
\ket{\zzm}\Bigr],
\end{split}
\end{align}
\ie, the Hilbert-space vectors $\zabk$ are the eigenvectors of
the matrix $\zmix$ given by
\begin{equation}
\label{eq:mixrho}
\zmix \defin \frac{1}{2}\zz 
+ \frac{1}{2} \zx 
= \frac{2 + \sqrt{2}}{4} \za
+ \frac{2 - \sqrt{2}}{4} \zb
\corr \frac{1}{4}\begin{pmatrix}
3&1\\1&1
\end{pmatrix},
\end{equation}
and they are orthogonal: $\tr(\za\zb)=0$. The action of the first
\diaphragm\ is thus
\begin{subequations}\label{eq:modifsepare}
\begin{align}
\zz &\mapsto 
\za,\text{ let through, with probability }
\tr(\za\zz) \approx 0.854,
\\
\zz &\mapsto 
\zb, \text{ reflected, with probability }
\tr(\zb\zz) \approx 0.146,
\\
\zx &\mapsto 
\za, \text{ let through, 
with probability $\tr(\za\zx)\approx 0.854$},\\
\zx &\mapsto 
\zb, \text{ reflected, with probability $\tr(\zb\zx)\approx
0.146$},
\end{align}
\end{subequations}
while for the second \diaphragm:
\begin{subequations}\label{eq:modifseparetwo}
\begin{align}
\zz &\mapsto 
\za,\text{ reflected, with probability }
\tr(\za\zz) \approx 0.854,
\\
\zz &\mapsto 
\zb, \text{ let through, with probability }
\tr(\zb\zz) \approx 0.146,
\\
\zx &\mapsto 
\za, \text{ reflected, 
with probability $\tr(\za\zx)\approx 0.854$},\\
\zx &\mapsto 
\zb, \text{ let through, with probability $\tr(\zb\zx)\approx
0.146$}.
\end{align}
\end{subequations}

\begin{figure}[t]
\setlength{\unitlength}{0.0047\columnwidth}
\begin{picture}(140,50)(0,0)
\put(0,0){\framebox(55,40){\scriptsize$\begin{aligned}&0.5\,\ketbra{\zzp}{\zzp}+\\&0.5\,\ketbra{\zxp}{\zxp}\end{aligned}$}}
\put(0,41){\makebox(0,0)[bl]{\scriptsize(a)}}

\put(57,20){\vector(1,0){18.5}}
\put(66.25,21){\makebox(0,0)[b]{\scriptsize$Q<0$}}

\put(77.5,6){\framebox(55,34){\scriptsize$\ketbra{\zap}{\zap}$}}
\put(77.5,0){\framebox(55,6){\scriptsize$\ketbra{\zbp}{\zbp}$}}
\put(77.5,41){\makebox(0,0)[bl]{\scriptsize(b)}}
\end{picture}
\caption{Separation of non-one-shot-distinguishable quantum gases.}\label{fig:nondistin}
\end{figure}
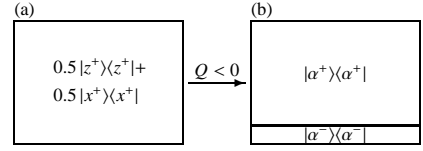

By inserting these \diaphragm s in the container as in the
preceding example, they will \emph{transform} and separate into
two chambers our $\zz$- and $\zx$-gases, leaving in the end an
$\za$-gas in the upper chamber and an $\zb$-gas in the lower one
(Fig.~\ref{fig:nondistin}). The gases will have the same pressure
as the original ones but occupy unequal volumes, because the
ratio for the $\zz$- and $\zx$-gases to be transformed into an
$\za$-gas and an $\zb$-gas was approximately $0.854/0.146$, as
seen from \eqns~\eqref{eq:modifsepare}; this will also be the
ratio of the final volumes. The total amount of work we spent for
this separation is
\begin{multline}\label{eq:worknoondingsepar}
-0.146\, \zN \zk \zT \ln \frac{0.146\, V}{V} - 0.854\, \zN \zk \zT
\ln \frac{0.854\, V}{V} \approx{}\\ 0.416\, \zN \zk \zT
\end{multline}
(the first term is for the upper \diaphragm, opaque to the
$\zb$-gas, and the second for the lower \diaphragm, opaque
to the $\za$-gas). This is also the amount of heat released by
the gases. 

The significance of the equations above is that the half/half
mixture of $\zz$- and $\zx$-gases can be treated as a
$0.854/0.146$ mixture of an $\za$-gas and an $\zb$-gas. In fact,
the two semi-permeable \diaphragm s could be used to completely
separate an $\za$-gas from an $\zb$-gas, whose preparations are
one-shot distinguishable.


\medskip

We shall see an application of the coupling between the
thermomechanic and quantum degrees of freedom provided by
semi-permeable \diaphragm s in the second part of the paper. For
the moment, we want to stress the following two-way connexion:
the existence of a measurement which can distinguish two
preparations in one shot implies the possibility of constructing
a semi-permeable \diaphragm\ which can distinguish and separate
the two preparations. \latin{Vice versa}, if we had such a
\diaphragm\ we could use it as a measurement device to
distinguish in one shot the two preparations, and so it would
imply their one-shot distinguishability.

The notion of a semi-permeable \diaphragm\ implies of course
something more, \viz, the possibility of a coupling between the
quantum degrees of freedom to be distinguished and the
translational degrees of freedom. It is this coupling, done with
a hybrid quantum-classical framework
(von~Neumann~\citep{vonneumann1932c_r1996} and the present paper)
or completely within quantum mechanics (Peres~\citep{peres1995}),
that allows applications and consequences of thermodynamic
character. However, the two-way connexion pointed out above is
independent of the fact that the \diaphragm\ may \emph{also} have
applications or consequences of thermodynamic character.

\subsection{Distinguishability, orthogonality,
and the second law}
\label{sec:distorthsecondlaw}

We can now summarise the observations and results that we have
gathered up to now: 

The existence of a measurement which can distinguish two
preparation procedures in one shot is equivalent to the existence
of semi-permeable \diaphragm s which can separate the two
preparations, and is also equivalent, by definition, to saying
that the two preparations are one-shot distinguishable.
Mathematically, the measurement and the \diaphragm s are then
represented by a \POVM\ satisfying \eqns~\eqref{eq:disting}, or
equivalently by the orthogonality of the density matrices
representing the two preparations.

Let us now look again at the two statements~\eqref{eq:first}
and~\eqref{eq:second}. We understand that, by ``state'',
von~Neumann meant a density matrix (or a Hilbert-space ray, which
can be considered as a particular case thereof).

The first statement~\eqref{eq:first},
\begin{quote}
``Two [density matrices] $\zphi$, $\zpsi$ can certainly be
separated by a semi-permeable \diaphragm\ if they are
orthogonal'',
\end{quote}
seems to be saying just this: the orthogonality of two density
matrices (\emph{represents} (mathematically) the fact that their
respective preparation procedures are one-shot distinguishable,
\ie, can be separated by some device. The demonstration of this
statement does not really require thermodynamic arguments, as
we have seen.\footnote{What von~Neumann's demonstration of the
first statement really shows is something else: \viz, that his
quantum entropy expression is consistent with thermodynamics.
This, however, does not concern us in the present paper, and will
perhaps be analysed elsewhere~\citep{bjoerketal2004b}.}

The second statement was:
\begin{quote}
``If [the density matrices] $\zphi$, $\zpsi$ are not orthogonal,
then the assumption of such a semi-permeable \diaphragm\ [which
can separate them] contradicts the second law of
thermodynamics''.
\end{quote}
Let us now try to analyse it. The statement mix together
mathematical concepts (density matrices, non-orthogonality) and
physical ones (semi-permeable \diaphragm), so we may try to state
it either in physical or in mathematical terms only.

Saying that two density matrices are not orthogonal is the
mathematical way of saying that their corresponding preparation
procedures are not one-shot distinguishable. So the statement
becomes:
\begin{quote}
If two preparations are not one-shot distinguishable, then the
assumption of a semi-permeable \diaphragm\ which can separate
them contradicts the second law.
\end{quote}
But in the previous sections we have shown that assuming the
existence of a separating semi-permeable \diaphragm\ is
equivalent by construction to assuming the existence of a
measurement which can distinguish the preparations in one shot.
So what the second statement is saying is just the following:
\begin{quote}
If two preparations are \emph{not} one-shot distinguishable,
then the assumption that they \emph{are} one-shot distinguishable
contradicts the second law.
\end{quote}

So the statement assumes that two preparations are one-shot
distinguishable \emph{and} not one-shot distinguishable; or in
mathematical terms, that their density matrices are orthogonal
\emph{and} not orthogonal. It has the logical form `$(A \land
\lnot A) \lmatimplies B$', and makes thus little sense, because
it starts from contradictory premises. In particular, it cannot
have any physical implications for the second law of
thermodynamics. We must in fact remember that, in a logical
formal system, from contradictory assumptions (`$A \land \lnot
A$') one can idly deduce any proposition whatever (`$B$') as well
as its negation (`$\lnot B$') \citete{see
\eg~\citeap{copietal1953_r1990,copi1954_r1979,hamilton1978_r1993}}.

The same contradictory premises are also present in Peres'
demonstration of the second statement by means of a
thought-experiment. This experiment will be discussed in detail
in the second part of this paper, for a different purpose. But we
can anticipate its basic idea only in order to see where the
contradictory premises lie:

A container is divided into two chambers containing two
quantum-ideal-gas samples. These are represented by
\emph{non-orthogonal} density matrices. Two semi-permeable
\diaphragm s are then used to \emph{completely separate} the two
samples, isothermally, just as in the first example of
\sect~\ref{sec:matintro}. In this way some heat is absorbed by
the gases, or equivalently, some work is done by them. The
process is then completed so as to come back to the initial
situation, and it is easily shown that a net amount of heat is
absorbed by the gases (or a net amount of work done by them),
violating the second law.

The contradictory premises are the following. It is assumed, on
the one hand, that the two initial preparations of the
quantum-gas samples are \emph{not} one-shot distinguishable, as
is reflected by the use of non-orthogonal density matrices; \ie,
that there does \emph{not} exist a measurement procedure able to
distinguish and separate them in one shot. On the other hand, it
is assumed immediately thereafter that there \emph{is} such a
measurement procedure, as is reflected by the existence of the
separating \diaphragm s which must necessarily implement it; in
particular, the quantum-gases have then to be represented by
\emph{orthogonal} density matrices. This is a contradiction of
course, and it has nothing to do with thermodynamic, work
extraction, or heat absorption. The thought-experiment cannot be
continued if we do not state clearly which of the mutually
exclusive alternatives, one-shot distinguishability or
not-one-shot distinguishability, \ie\ orthogonality or
non-orthogonality, is the one meant to hold.


Note that we are \emph{not} saying that there are \emph{no}
relationships or valid consistency considerations between quantum
mechanics and thermodynamics. In fact, what many of von~Neumann's
and Peres' analyses and thought-experiments really demonstrate is
that the quantum-mechanical physical concepts and principles are
consistent with the thermodynamic ones, which is a fundamental
result. As Peres~\citep[p.~275]{peres1995} states it, ``It thus
appears that thermodynamics imposes severe constraints on the
choice of fundamental axioms for quantum
theory''.\footnote{\label{fn:reviewer2}We thank an anonymous
reviewer for the \emph{American Journal of Physics} 
for pointing out this passage to us. In the same paragraph, Peres
also mention a \emph{petitio principii} which, however, concerns
the proof of the equivalence of the von~Neumann and thermodynamic
entropies.} What we are saying is that the existence of a
semi-permeable \diaphragm\ that can separate two preparations,
and the non-one-shot distinguishability of these preparations,
are contradictory assumptions \emph{within} the
quantum-mechanical theory itself, and so cannot be simultaneously
used to study the consistency of this theory with another one
(thermodynamics). Using an example given in
\sect~\ref{sec:lingdet}: we can discuss the consistency of a
particular microscopic force field with macroscopic thermodynamic
laws; but it is vain to assume that the force field is
conservative and its integral along a closed path does not vanish
in order to derive a violation of the second law --- these
assumptions are contradictory by definition and anything can be
vacuously derived from them.

Peres' thought-experiment, though, does not lose its value
because of its contradictory premises. The reason lies in its
subtle (and nice) presentation: not as an impersonal reasoning,
but as an \emph{interplay} between \emph{two observers} (us and a
``wily inventor'', as Peres calls him~\citep[p.~275]{peres1995}).
It is thus possible to re-interpret the contradictory premises as
\emph{different, contrasting pieces of knowledge} of two
observers. This will be now done in the second part of the paper.

\section{Part II: Contrasting pieces of knowledge and violations
of the second law}
\label{part:II}
\begin{flushright}
{\footnotesize\begin{minipage}{.7\columnwidth}
{\footnotesize\emph{``\emph{One} can't perhaps,'' said Humpty
Dumpty; ``but \emph{two} can. With proper assistance, you might
have left off at seven.''}~\citep{carroll1871}}
\end{minipage}\par}
\end{flushright}

\subsection{The dependence of a density matrix on the observer's
knowledge and on the set of preparations and measurements chosen}
\label{sec:depdensmatr}

The fact that a density matrix always depends on the particular
knowledge about the properties of the preparation it represents,
has been stressed amongst others by
Kemble~\citep[p.~1021]{kemble1939}\citep[pp.~1155--57]{kemble1939b},
Jaynes~\citep{jaynes1979b,jaynes1993b,jaynes1989,jaynes1990b,jaynes2003},
Caves
\etal~\citep{fuchs2001,fuchsetal2004,cavesetal2002b,cavesetal2002c},
and
ourselves~\citep{mana2003,mana2004b,mana2005}.\footnote{Caves,
Fuchs, and Schack see the density matrix as a sort of ``quantum''
analogue of a probability, and develop a ``quantum'' probability
theory in analogy with (``classical'') probability theory (with
analogous theorems, like \eg\ Bayes' and
de~Finetti's~\citep{fuchs2001,schacketal2001,cavesetal2002,cavesetal2002b,fuchsetal2004,fuchsetal2004b,definetti1937b}).
We should instead prefer to derive the quantum formalism as a
\emph{particular} application of probability
theory~\citep{mana2004b,mana2005}.}
Discussions and analyses of the compatibility of density-matrix
assignments by different observers have already been offered,
\eg, by Brun \etal~\citep{brunetal2002}, and Caves
\etal~\citep{cavesetal2002c}. 

Note in particular that a density matrix can be associated to a
preparation (and a \POVM\ to a measurement) only \emph{after} we
have specified the \emph{whole} sets of preparation and
measurement procedures we want or have to consider --- indeed, it
is quite appropriate to say that these sets \emph{define} our
`system'. This might appear paradoxical to some, but is clearly
reflected in the fact that, in the description of a phenomenon,
we always have to choose a particular Hilbert space before
introducing any density matrix.
This fact can be inferred from Peres'~\citep{peres1995} and
Hardy's~\citep{hardy2001} work, and we have tried to make it
mathematically explicit
elsewhere~\citep{mana2003,mana2004b,mana2005}.

Thus, if we add a single measurement procedure --- perhaps a
newly discovered one --- to the chosen set of measurement
procedures, we must in some cases change (numerically and even
dimensionally) the density matrices we had previously associated
to the preparations.\footnote{This fact is related to the
restrictions (``compatibility'' or ``positivity domains'') of the
set of statistical matrices for the ``reduced dynamics'' of some
open quantum
systems~\citep{pechukas1994,stelmachovicetal2001,hayashietal2003,fonsecaromeroetal2003,jordanetal2004,jordanetal2005,jordan2005}.
These restrictions simply arise because one wants to describe a
set of preparation procedures by means of a density-matrix space
that is instead only meant for, and is only appropriate for, a
particular smaller set. On the other hand, the reason why one
wants to do this is that quantum mechanics has the following
unfortunate property: if you want to add just one more
preparation or measurement procedure to the sets that interest
you, you must be ready to ``pay'' for at least $2N+1$ more
dimensions for your matrix-spaces, where $N^2-1$ is the dimension
of the (normalised-Hermitean-matrix) spaces you were using (\ie,
you pay the difference between $(N+1)^2-1$ and $N^2-1$). Compare
this with the classical ``cost'' of only $1$ more dimension.}

\subsection{The  value of Peres' thought-experiment from a
different perspective}
\label{sec:valueperes}

Peres' thought-experiment is particularly suited to show the
points above. In its presentation, two observers have contrasting
pieces of knowledge about the preparations of the quantum ideal
gases: the first observer thinks them not to be one-shot
distinguishable, and represents them accordingly by
non-orthogonal density matrices; the second observer, instead,
knows they are one-shot distinguishable and, in fact, he
possesses actual devices --- the semi-permeable \diaphragm s ---
by which he separates them. This is reflected in the fact that
the set of measurement procedures considered by the second
observer contains (at least) a measurement (\viz, the one which
can distinguish in one shot the gas preparations) not contained
in the set of the first observer, simply because the first
observer did not know about the existence of that measurement.

A similar, two-observer style of presentation was chosen by
Jaynes~\citey[\sect~6]{jaynes1992} for a thought-experiment which
showed that two different observers may have different pieces of
knowledge and so give different thermodynamic descriptions for
the same physical situation. When they interact, strange results
may arise; \eg, it may appear to one of them as if the other were
violating the second law. From this point of view Jaynes' and
Peres' thought-experiments are very similar also in their
results; the similarity is the more fascinating because Jaynes'
experiment is completely within a classical, not quantum,
context.

Grad~\citep[p.~325]{grad1961} also remarked that not only two
observers, but even more simply the same observer facing two
different applications, may use different pieces of knowledge and
descriptions to study the same physical situation.

For this reason, we shall analyse Peres' thought-experiment with
the following scheme. First, we present it from the point of view
of the first observer, that we call Tatiana, which sees a seeming
violation of the second law. Then we go over to Jaynes' simpler
thought-experiment, and see how the two observers' (Johann and
Marie) contrasting descriptions are resolved there. With the
insight provided by Jaynes' experiment, we turn back to Peres',
and re-analyse it from the point of view of the second observer,
that we call Willard, seeing that he had some additional
knowledge and measurement means with respect to Tatiana, and that
for him the second law is not violated; on the other hand,
Tatiana will have to change the dimensions and numerical values
of the density matrices used in her description if she wants to
make allowance for Willard's new measurement
procedure.\footnote{In some situations it might perhaps be
necessary to abandon the quantum-theoretical physical principles
partially or altogether, in favour of other more economical or
aesthetically pleasing. Note that it is easily
proven~\citep{mana2005} (\cf~\citep[\sect~V]{mana2003},
Holevo~\citep[\chap~I]{holevo1980_t1982}) that there always exist
appropriate (if necessary, infinite-dimensional)
quantum-mechanical density-matrix- and \POVM-spaces by which one
can represent sets of preparation and measurement procedures
having \emph{any} statistical properties whatever. In this sense,
the quantum-theoretical \emph{mathematical} formalism can never
be ``proven wrong'' (but note that the same also holds for the
classical statistical
formalism~\citep[\sect~I.7]{holevo1980_t1982}), although it may
be redundant (see preceding footnote).}

\subsection{Peres' thought-experiment: Tatiana's description\label{sec:peresdem}}

Peres' demonstration~\citep[pp.~275--277]{peres1995} can be
presented as follows. We are in a quantum laboratory, where the
physicist Tatiana is studying a container wherein two
quantum-ideal-gas samples are confined to two chambers, having
volume $V/2$ each and separated by an impermeable \diaphragm.
From some measurements made on a series of identically prepared
containers, she has chosen to describe the internal quantum
degrees of freedom of the ideal gases by means of the density-\bd
matrix space for a quantum two-level system. In her description,
the upper chamber contains a $\zz$-gas, the lower an $\zx$-gas,
where $\zz$ and $\zx$ are the density matrices defined in
\sect~\ref{sec:matintro}, \eqns~\eqref{eq:zplus}
and~\eqref{eq:xplus}. They are non-\bd orthogonal, $\tr(\zz\zx)
\neq 0$, because to Tatiana's knowledge there are no means to
distinguish the two corresponding preparations in one shot, as
she could not separate the two gases completely with any
semi-permeable \diaphragm s known to her.

Enters a ``wily inventor''; let us call him Willard. He claims
having produced two such semi-permeable \diaphragm s, which can
completely separate the two gases.\footnote{This being a
scientific paper, we ideally (\ie, unrealistically) assume
Willard's honesty and the absence of any fraud.} In fact, by
means of them he reversibly mixes the two gases, \emph{obtaining}
work equal to $Q'=\zN \zk \zT \ln 2 \approx 0.693\, \zN
\zk \zT$ (
\cf\ \eqn~\eqref{eq:worksepar}). From Tatiana's point of view
this is quite surprising!

\begin{figure}[t]
\setlength{\unitlength}{0.0047\columnwidth}
\begin{picture}(210,109)(0,-60)
\put(0,20){\framebox(55,20){\scriptsize$\ketbra{\zzp}{\zzp}$}}
\put(0,0){\framebox(55,20){\scriptsize$\ketbra{\zxp}{\zxp}$}}
\put(0,41){\makebox(0,0)[bl]{\scriptsize(a)}}

\put(57,20){\vector(1,0){18.5}}
\put(66.25,21){\makebox(0,0)[b]{\scriptsize$Q'>0$}}
\put(66.25,19){\makebox(0,0)[t]{\scriptsize?}}

\put(77.5,0){\framebox(55,40){\scriptsize$\begin{aligned}&0.5\,\ketbra{\zzp}{\zzp}+\\&0.5\,\ketbra{\zxp}{\zxp}\end{aligned}$}}
\put(77.5,41){\makebox(0,0)[bl]{\scriptsize(b)}}

\put(134.5,20){\vector(1,0){18.5}}
\put(143.75,21){\makebox(0,0)[b]{\scriptsize$\equiv$}}

\put(155,0){\framebox(55,40){\scriptsize$\begin{aligned}&0.85\,\ketbra{\zap}{\zap}+\\&0.15\,\ketbra{\zbp}{\zbp}\end{aligned}$}}
\put(155,41){\makebox(0,0)[bl]{\scriptsize(c)}}

\put(182.5,-2){\vector(0,-1){18.5}}
\put(183.5,-11.25){\makebox(0,0)[l]{\scriptsize$-Q''<Q'$}}

\put(155,-62.5){\framebox(55,6){\scriptsize$\ketbra{\zbp}{\zbp}$}}
\put(155,-56.5){\framebox(55,34){\scriptsize$\ketbra{\zap}{\zap}$}}
\put(155,-21.5){\makebox(0,0)[bl]{\scriptsize(d)}}


\put(153,-42.5){\vector(-1,0){18.5}}

\put(77.5,-42.5){\framebox(55,20){\scriptsize$\ketbra{\zzp}{\zzp}$}}
\put(77.5,-62.5){\framebox(55,20){\scriptsize$\ketbra{\zzp}{\zzp}$}}
\put(77.5,-21.5){\makebox(0,0)[bl]{\scriptsize(e)}}

\put(75.5,-42.5){\vector(-1,0){18.5}}

\put(0,-42.5){\framebox(55,20){\scriptsize$\ketbra{\zzp}{\zzp}$}}
\put(0,-62.5){\framebox(55,20){\scriptsize$\ketbra{\zxp}{\zxp}$}}
\put(0,-21.5){\makebox(0,0)[bl]{\scriptsize(f)${}\equiv{}$(a)}}

\put(27.5,-20.5){\vector(0,1){18.5}}
\put(26.5,-11.25){\makebox(0,0)[r]{\scriptsize${}\equiv{}$}}



\end{picture}
\caption{Quantum gas experiment from Tatiana's point 
of view.}\label{fig:alfred}
\end{figure}
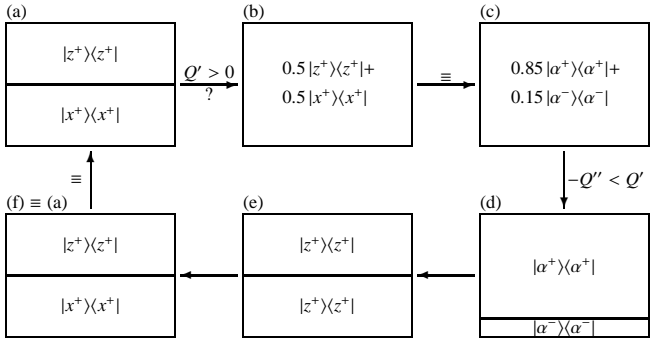

Now she has a single container of volume $V$ filled with a
half/half mixture of $\zz$- and $\zx$-gases
(Fig.~\ref{fig:alfred},~b). She wants to return to the initial
situation, but she cannot use Willard's puzzling means, of
course. For her the situation is now the same as that discussed
in the second example of \sect~\ref{sec:matintro}: the gas
mixture is for her equivalent (Fig.~\ref{fig:alfred},~d) to
a mixture of approximately $0.854$ parts of an $\za$-gas and
$0.146$ parts of an $\zb$-gas, where $\za$ and $\zb$ are the
density matrices defined in \eqn~\eqref{eq:newketalf}. Tatiana
uses two semi-permeable \diaphragm s to separate the two
$\zab$-gases, the $\za$-gas into a $0.854$ fraction of the volume
$V$, and the $\zb$-gas into the remaining $0.146$ fraction (so
that they have the same pressure), and \emph{spends} work equal
to $-Q'' = -\zN \zk \zT (0.854 \ln 0.854 + 0.146 \ln
0.146)\approx 0.416\, \zN \zk \zT$, \cf\ 
\eqn~\eqref{eq:worknoondingsepar}.

Tatiana then performs two operations corresponding to unitary
rotations which reversibly change the density matrices associated
to the two gases into the same density matrix, say $\zz$, so that
the two chambers eventually contain for her the same $\zz$-gas.
She then eliminates the \diaphragm s and reinserts another
impermeable one in the middle to divide the gas into two chambers
of equal volume (Fig.~\ref{fig:alfred},~e) --- which is, from her
point of view, a reversible operation ---, and finally performs
again an operation represented by a rotation $\zz \mapsto \zx$ of
the density matrix associated to the gas in the lower chamber. In
this way she has apparently re-established the original condition
of the gases (Fig.~\ref{fig:alfred},~a), which have thus undergone
a cycle. The last operations were assumed to be performable
without expenditure or gain of work, hence without heat exchange
either.%
\footnote{Von~Neumann~\citep[pp.~194, 197]{vonneumann1932c_r1996}
and Peres~\citep[p.~275]{peres1995} assert that unitary rotations
can be realised without \emph{heat} exchange, but work exchange
is allowed and indeed sometimes necessary. However, in our
present discussion we have assumed all processes to be isothermal
and all gases ideal, which implies that any reversible
\emph{isochoric} work exchange (like the unitary rotations) would
be accompanied by an equivalent reversible heat exchange (see
\sect~\ref{sec:matintro}), with a consequent undesired entropy
change. This is why Tatiana's final isochoric unitary rotations
must be performed with no energy exchange. This issue is related
to the problematic way in which the quantum and classical or
thermodynamic descriptions are combined; namely, the density
matrices are not thermodynamic variables \langfrench{au pair}
with the real numbers ($V$ and $T$) describing the gas; \cf\ 
\sect~\ref{sec:matintro}.}

Tatiana summarises the results as follows: a cycle has been
completed because the initial and final situations are the same.
The total heat $Q$ absorbed by the gases equals the work ---
experimentally measured --- done by them and amounts to
\begin{equation}
\label{eq:heatabscycle}
Q =Q' + Q'' \approx (0.693 - 0.416)\, \zN \zk \zT = 
0.277\, \zN \zk \zT > 0.
\end{equation}
Hence, she sees a violation of the second
law~\eqref{eq:seclawcycl} because, from her point of view,
\begin{equation}
\label{eq:viollaw}
Q > 0\quad\text{in the cyclic process}.
\end{equation}

Tatiana accuses Willard of having violated the second law by
means of his strange semi-permeable \diaphragm s that
``completely separate non-\bd orthogonal density matrices''.

\medskip

Has the second law really been violated? Also from Willard's
point of view? We do not answer these questions now; instead, we
leave the quantum laboratory where Tatiana and Willard are now
arguing after their experiment, and enter an adjacent classical
laboratory, where we shall look at Jaynes'
demonstration~\citep{jaynes1992}. The situation there is in many
respects very similar to the previous one, though it is
\emph{completely} ``classical''.

\subsection{Jaynes' thought-experiment\label{sec:jaynesdem}}

\begin{figure}[t]
\setlength{\unitlength}{0.0047\columnwidth}
\begin{picture}(210,57)(0,-8)
\put(0,20){\framebox(55,20){\scriptsize Ar}}
\put(0,0){\framebox(55,20){\scriptsize Ar}}
\put(0,41){\makebox(0,0)[bl]{\scriptsize(a)}}

\put(57,20){\vector(1,0){18.5}}
\put(66.25,21){\makebox(0,0)[b]{\scriptsize$Q'>0$}}
\put(66.25,19){\makebox(0,0)[t]{\scriptsize?}}

\put(77.5,0){\framebox(55,40){\scriptsize Ar}}
\put(77.5,41){\makebox(0,0)[bl]{\scriptsize(b)}}

\put(134.5,20){\vector(1,0){18.5}}

\put(155,20){\framebox(55,20){\scriptsize Ar}}
\put(155,0){\framebox(55,20){\scriptsize Ar}}
\put(155,41){\makebox(0,0)[bl]{\scriptsize(c)${}\equiv{}$(a)}}

\put(182.5,-2){\line(0,-1){6}}
\put(182.5,-8){\line(-1,0){155}}
\put(27.5,-8){\vector(0,1){6}}
\put(105,-7){\makebox(0,0)[b]{\scriptsize${}\equiv{}$}}
\end{picture}
\caption{Classical gas experiment from Johann's point 
of view.}\label{fig:johann}
\end{figure}
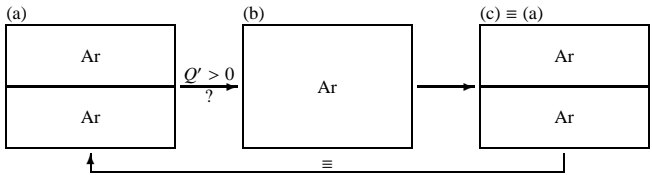

In the classical laboratory, we have a container wherein an
ideal-gas sample is equally divided into two chambers of volumes
$V/2$ each and separated by an impermeable \diaphragm\ 
(Fig.~\ref{fig:johann},~a). From the measurements made by the
scientist Johann, the gas in the two chambers is exactly the
same, ``ideal argon''.\footnote{Real argon, of course, behaves
like an ideal gas only in certain ranges of temperature and
volume.} For Johann it would thus be impossible, not to say
meaningless, to find a semi-permeable \diaphragm\ that be
transparent to the gas in the upper chamber and opaque to the gas
in the lower one, and another \diaphragm\ with the opposite
properties.

The scientist Marie, also in the laboratory, states nevertheless
that she has in fact two \diaphragm s with those very properties.
She uses them to reversibly and isothermally mix the two halves
of the gas, \emph{obtaining} work equal to $Q' = \zN \zk \zT \ln
2 \approx 0.693\, \zN \zk \zT$ (Fig.~\ref{fig:johann},~b), and
leaving Johann stupefied.

In fact, from Johann's point of view Marie has left things
exactly as they were: he just needs to reinsert the impermeable
\diaphragm\ in the middle of the container and for him the
situation is exactly the same as in the beginning: the \emph{one}
gas is equally divided into two equal chambers
(Fig.~\ref{fig:johann},~c~a).

Johann's conclusion is the following: The initial and final
conditions of the gas are the same, so a cyclic process has been
completed. The work obtained --- experimentally measured ---
equals the heat absorbed by the gas,
\begin{equation}
\label{eq:heatabscyclejaynes}
Q = Q' \approx 0.693\, \zN \zk \zT > 0,
\end{equation}
and so for Johann
\begin{equation}
Q  > 0\quad\text{in the cyclic process}
\end{equation}
(\cf\ Tatiana's \eqn~\eqref{eq:viollaw}), which plainly
contradicts the second law of
thermodynamics~\eqref{eq:seclawcycl}.

\medskip

We see that what happens here is completely analogous to what has
happened in the quantum laboratory: a physicist has, from initial
measurements, a particular description of a given situation. Then
a process takes place that contradicts the physicist's
mathematical description, and the re-establishment of what was
thought to be the initial situation yields an apparent
experimental violation of the second law.

The reader has no doubt noticed that the facts were presented not
only from Johann's point of view, but also, so to speak, taking
side with him. In fact, we plainly ignored Marie's experimental
performance in our mathematical description. But it should be
clear that, since the two gas samples in the chambers behaved
differently with respect to Marie's semi-permeable \diaphragm s
(one sample exerted pressure on one of the \diaphragm s but not
on the other, and \latin{vice versa} for the other sample), then
they must actually be samples of \emph{two different}
gases,\footnote{Provided, of course, that this phenomenon is also
reproducible.} contrary to what Johann (and we) believed and
described mathematically.

Note that this fact does not completely contradicts Johann's
point of view. It simply means that the two samples behave
exactly in the same manner with respect to Johann's experimental
and measurement means, and so he was justified to consider them
as samples of the same gas. But now Johann has experimental
evidence that the two samples behave differently in certain
circumstances, and so, in order to avoid inconsistencies, they
\emph{have to} be considered as samples of different gases, at
least in all experimental situations in which their difference in
behaviour comes about (such as the mixing performed by Marie).

So let us see how the whole process has taken place from Marie's
point of view. She explains that the gas samples initially
contained in the two chambers were two different kinds of
ideal-argon, of which Johann had no knowledge: `argon~$a$'
($\zAa$) and `argon~$b$' ($\zAb$). Argon~$a$ is soluble in
whafnium while argon~$b$ is not, but the latter is soluble in
whifnium, a property not shared by the
$a$~variety.\footnote{Jaynes~\citep[\sect~5]{jaynes1992} explains
that `whifnium', as well as `whafnium', ``is one of the rare
superkalic elements; in fact, it is so rare that it has not yet
been discovered''.} Marie's separation of the two gases $\zAa$
and $\zAb$ was possible by means of two semi-permeable \diaphragm
s made of whifnium and whafnium that take advantage of these
different properties. (Of course, Jaynes' and our speaking of
`argon~$a$', `argon~$b$', `whafnium', and `whifnium' in this
imaginary experiment is just a coloured way of stating that the
two samples are of different gases. The reader can, if he or she
so prefers, simply call them `gas $A$' and `gas $B$' and take
into account their different behaviour with respect to the two
semi-permeable \diaphragm s.)

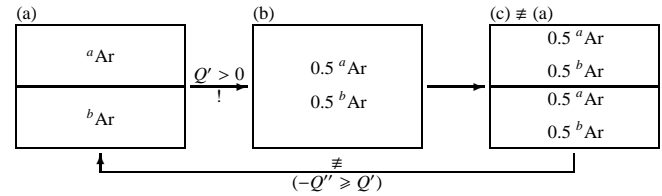
\begin{figure}[t]
\setlength{\unitlength}{0.0047\columnwidth}
\begin{picture}(210,57)(0,-8)
\put(0,20){\framebox(55,20){\scriptsize${}^a\text{Ar}$}}
\put(0,0){\framebox(55,20){\scriptsize${}^b\text{Ar}$}}
\put(0,41){\makebox(0,0)[bl]{\scriptsize(a)}}

\put(57,20){\vector(1,0){18.5}}
\put(66.25,21){\makebox(0,0)[b]{\scriptsize$Q'>0$}}
\put(66.25,19){\makebox(0,0)[t]{\scriptsize!}}

\put(77.5,0){\framebox(55,40){\scriptsize$\begin{aligned}&0.5\;{}^a\text{Ar}\\&0.5\;{}^b\text{Ar}\end{aligned}$}}
\put(77.5,41){\makebox(0,0)[bl]{\scriptsize(b)}}

\put(134.5,20){\vector(1,0){18.5}}

\put(155,20){\framebox(55,20){\scriptsize$\begin{aligned}&0.5\;{}^a\text{Ar}\\&0.5\;{}^b\text{Ar}\end{aligned}$}}
\put(155,0){\framebox(55,20){\scriptsize$\begin{aligned}&0.5\;{}^a\text{Ar}\\&0.5\;{}^b\text{Ar}\end{aligned}$}}
\put(155,41){\makebox(0,0)[bl]{\scriptsize(c)${}\nequiv{}$(a)}}

\put(182.5,-2){\line(0,-1){6}}
\put(182.5,-8){\line(-1,0){155}}
\put(27.5,-8){\vector(0,1){6}}
\put(105,-7){\makebox(0,0)[b]{\scriptsize${}\nequiv{}$}}
\put(105,-9){\makebox(0,0)[t]{\scriptsize($-Q'' \ge Q'$)}}

\end{picture}
\caption{Classical gas experiment from Marie's point 
of view.}\label{fig:marie}
\end{figure}

According to Marie (Fig.~\ref{fig:marie}), the second law is
\emph{not} violated: Initially the two gases $\zAa$ and $\zAb$
were completely separated in the container's two chambers
(Fig.~\ref{fig:marie},~a). After her mixing and extracting work,
the container contained an equal mixture of $\zAa$ and $\zAb$
(Fig.~\ref{fig:marie},~b). Upon Johann's reinsertion of the
impermeable \diaphragm\ the container was again divided in two
equal chambers, but each chamber contained a half/half mixture of
$\zAa$ and $\zAb$ (Fig.~\ref{fig:marie},~c), and this was
\emph{different} from the initial condition
(Fig.~\ref{fig:marie},~a). Thus \emph{the cycle was not
completed, although it appeared so to Johann}, and so the form of
the second law for cyclic processes, \eqn~\eqref{eq:seclawcycl},
cannot be applied. To close the cycle one has to use the
semi-permeable \diaphragm s again to relegate the two gases to
two separate chambers, and must thereby \emph{spend} an amount of
work $-Q''$ at least equal to that previously obtained, $-Q''\ge
Q'$, and the second law~\eqref{eq:seclawcycl} for the completed
cycle is satisfied: $Q = Q' + Q'' \le 0$.

The simple conclusion, stated in terms of entropy, that
Jaynes~\citep[\sect~3]{jaynes1992} draws from this demonstration,
is that
\begin{quote}
it is necessary to decide at the outset of a problem which
macroscopic variables or degrees of freedom we shall measure
and/or control; and within the context of the thermodynamic
system thus defined, entropy will be some function $S(X_1,
\dotsc, X_n)$ of whatever variables we have chosen. We can expect
this to obey the second law [$Q/\zT \le \incr S$] only as long as
all experimental manipulations are confined to that chosen set.
If someone, unknown to us, were to vary a macrovariable $X_{n+1}$
outside that set, he could produce what would appear to us as a
violation of the second law, since our entropy function $S(X_1,
\dotsc, X_n)$ might decrease spontaneously [\ie, without
absorption of heat ($Q = 0$)], while his $S(X_1, \dotsc, X_n,
X_{n+1})$ increases.
\end{quote}

This is old wisdom: Grad~\citep[p.~325]{grad1961} (see
also~\citeap{grad1952,grad1967}) explained thirty-one years
earlier that
\begin{quote}
the adoption of a new entropy is forced by the discovery
of new information. [\ldots] The existence of diffusion
between oxygen and nitrogen somewhere in a wind tunnel
will usually be of no interest. Therefore the
aerodynamicist uses an entropy which does not recognise
the separate existence of the two elements but only that
of ``air''. In other circumstances, the possibility of
diffusion between elements with a much smaller mass ratio
(\eg, 238/235) may be considered quite relevant.
\end{quote}

But Jaynes' and Grad's remarks do not apply only to entropy; they
have greater generality. We always choose some variables --- with
particular ranges, scales, and governing equations --- to
describe a physical phenomenon. And such a choice always
represents, and is dependent on, the particular knowledge that we
have about that phenomenon, or that we think is sufficient to
describe it in a given situation or application.\footnote{For an
excellent discussion on different levels and scales of
description in the particular case of thermodynamics, see
Wood~\citey{woods1975}; also Samoh\'yl~\citep[especially
\sect~7]{samohyl1987} and Jaynes~\citep[\sect~1.2]{jaynes1965b}.}
This is true in particular for the quantum-mechanical density
matrices, Hilbert spaces, and so on.

\subsection{Peres' thought-experiment: Willard's description\label{sec:reanperes}}

With the insight provided by the analysis of the classical
experiment and by Grad's and Jaynes' remarks, we can return to
the quantum laboratory and look with different eyes at what
happened there.

Just as in the case of Johann and Marie, we must admit that in
the presentation of the quantum experiment we took not only
Tatiana's point of view, but also her parts, disregarding
Willard's experimental evidence in our mathematical description.
In fact, Willard showed that the two quantum ideal gases
\emph{can} be completely (and, it is assumed, reproducibly)
separated; and as shown in \sect~\ref{sec:distorthsecondlaw} this
means that there \emph{must} exist a measurement procedure by
which the two corresponding quantum preparations can be
distinguished in one shot. This is not in complete contradiction
with Tatiana's initial description: with the measurement means at
\emph{her} disposal, the two preparation procedures were not
distinguishable in one shot, and so \emph{for her} they were
appropriately represented by non-\bd orthogonal density matrices.
But, in order to avoid inconsistencies, the two preparation
procedures \emph{have to} be represented by orthogonal density
matrices in all experimental situations involving the new
measurement capability --- such as Willard's mixing process.

Willard thus represents the two quantum ideal gases by orthogonal
density matrices, and his mathematical analysis of the
thermodynamic process is different from Tatiana's. Let us follow,
step by step, a possible explanation of the quantum-gas process
from his point of view. He explains that the internal quantum
degrees of freedom of the gases are best represented by the
density-matrix space for a quantum four-level system (it might be
that the molecules of the gas where diatomic and not mono-atomic
as Tatiana believed), of which Tatiana only used a subspace
because of her limited measurement means (\eg, she probed the
internal quantum degrees of freedom of only one of the molecule's
atoms) . In other words, part of the density-matrix space used by
Willard was ``traced out'' in Tatiana's description, because she
had only access to measurement procedures represented by a
portion of the total \POVM\ space.

Denoting by $\set{%
\ket{\zzup}, \ket{\zzum}, \ket{\zzvp}, \ket{\zzvm}
}$ the orthonormal basis for the Hilbert space used by Willard,
Tatiana could not distinguish, amongst others, the preparations
corresponding to
$\ket{\zzup}$ and to $\ket{\zzvp}$, both of which she
represented as $\zzk$, nor those corresponding to
$\ket{\zxup} \defin \bigl(\ket{\zzup} +
\ket{\zzum}\bigr)/\sqrt{2}$ and to $\ket{\zxvp} \defin
\bigl(\ket{\zzvp} + \ket{\zzvm}\bigr)/\sqrt{2}$, which she
represented as $\zxk$. Tatiana's projection is thus of the kind
$\ket{\zphi \, \zpsi} \mapsto \ket{\zphi}$:
\begin{gather}
\label{eq:proj}
\begin{aligned}
\ket{\zzup}&\mapsto \zzk,&
\ket{\zzum}&\mapsto \ket{\zzm},\\
\ket{\zzvp}&\mapsto \zzk,&
\ket{\zzvm}&\mapsto \ket{\zzm},
\end{aligned}\\
\intertext{from which also follows}
\begin{aligned}
\ket{\zxup}&\mapsto \zxk,&
\ket{\zxum}&\mapsto \ket{\zxm},\\
\ket{\zxvp}&\mapsto \zxk,&
\ket{\zxvm}&\mapsto \ket{\zxm}.
\end{aligned}
\end{gather}

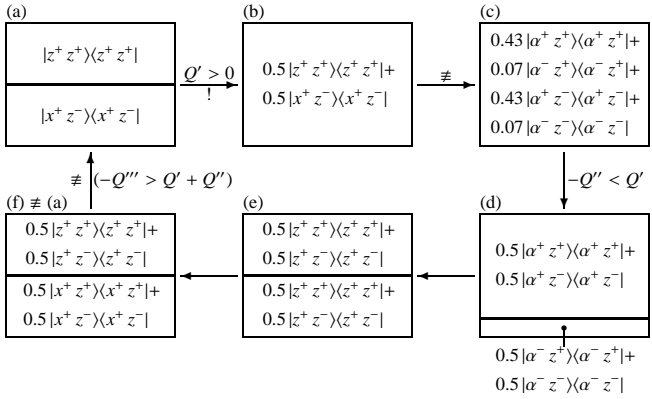
\begin{figure}[t]
\setlength{\unitlength}{0.0047\columnwidth}
\begin{picture}(210,129)(0,-82.5)
\put(0,20){\framebox(55,20){\scriptsize$\ketbra{\zzup}{\zzup}$}}
\put(0,0){\framebox(55,20){\scriptsize$\ketbra{\zxvp}{\zxvp}$}}
\put(0,41){\makebox(0,0)[bl]{\scriptsize(a)}}

\put(57,20){\vector(1,0){18.5}}
\put(66.25,21){\makebox(0,0)[b]{\scriptsize$Q'>0$}}
\put(66.25,19){\makebox(0,0)[t]{\scriptsize!}}

\put(77.5,0){\framebox(55,40){\scriptsize$\begin{aligned}&0.5\,\ketbra{\zzup}{\zzup}+\\&0.5\,\ketbra{\zxvp}{\zxvp}\end{aligned}$}}
\put(77.5,41){\makebox(0,0)[bl]{\scriptsize(b)}}

\put(134.5,20){\vector(1,0){18.5}}
\put(143.75,21){\makebox(0,0)[b]{\scriptsize$\nequiv$}}

\put(155,0){\framebox(55,40){\scriptsize$\begin{aligned}&0.43\,\ketbra{\zaup}{\zaup}+\\&0.07\,\ketbra{\zbup}{\zbup}+\\&0.43\,\ketbra{\zavp}{\zavp}+\\&0.07\,\ketbra{\zbvp}{\zbvp}\end{aligned}$}}
\put(155,41){\makebox(0,0)[bl]{\scriptsize(c)}}

\put(182.5,-2){\vector(0,-1){18.5}}
\put(183.5,-11.25){\makebox(0,0)[l]{\scriptsize$-Q''<Q'$}}

\put(155,-62.5){\framebox(55,6){}}
\put(155,-56.5){\framebox(55,34){\scriptsize$\begin{aligned}&0.5\,\ketbra{\zaup}{\zaup}+\\&0.5\,\ketbra{\zavp}{\zavp}\end{aligned}$}}
\put(155,-21.5){\makebox(0,0)[bl]{\scriptsize(d)}}

\put(182.5,-59.5){\line(0,-1){6}}\put(182.5,-59.5){\circle*{2}}
\put(182.5,-65.5){\makebox(0,0)[t]{\scriptsize$\begin{aligned}&0.5\,\ketbra{\zbup}{\zbup}+\\&0.5\,\ketbra{\zbvp}{\zbvp}\end{aligned}$}}

\put(153,-42.5){\vector(-1,0){18.5}}

\put(77.5,-42.5){\framebox(55,20){\scriptsize$\begin{aligned}&0.5\,\ketbra{\zzup}{\zzup}+\\&0.5\,\ketbra{\zzvp}{\zzvp}\end{aligned}$}}
\put(77.5,-62.5){\framebox(55,20){\scriptsize$\begin{aligned}&0.5\,\ketbra{\zzup}{\zzup}+\\&0.5\,\ketbra{\zzvp}{\zzvp}\end{aligned}$}}
\put(77.5,-21.5){\makebox(0,0)[bl]{\scriptsize(e)}}

\put(75.5,-42.5){\vector(-1,0){18.5}}

\put(0,-42.5){\framebox(55,20){\scriptsize$\begin{aligned}&0.5\,\ketbra{\zzup}{\zzup}+\\&0.5\,\ketbra{\zzvp}{\zzvp}\end{aligned}$}}
\put(0,-62.5){\framebox(55,20){\scriptsize$\begin{aligned}&0.5\,\ketbra{\zxup}{\zxup}+\\&0.5\,\ketbra{\zxvp}{\zxvp}\end{aligned}$}}
\put(0,-21.5){\makebox(0,0)[bl]{\scriptsize(f)${}\nequiv{}$(a)}}

\put(27.5,-20.5){\vector(0,1){18.5}}

\put(26.5,-11.25){\makebox(0,0)[r]{\scriptsize${}\nequiv{}$}}
\put(28.5,-11.25){\makebox(0,0)[l]{\scriptsize($-Q'''>Q'+Q''$)}}

\end{picture}
\caption{Quantum gas experiment from Willard's point 
of view.}\label{fig:willard}
\end{figure}

From Willard's point of view, the process went as follows. He had
also made some measurements on identically prepared containers,
and according to his results, the container's two chambers
initially contained $\zzu$- and $\zxv$-gases
(Fig.~\ref{fig:willard},~a), with
\begin{align}
\zzu &\defin \ketbra{\zzup}{\zzup} \corr
\left(\begin{smallmatrix}
1&0&0&0\\ 0&0&0&0\\ 0&0&0&0\\ 0&0&0&0
\end{smallmatrix}\right),\\
\zxv &\defin \ketbra{\zxvp}{\zxvp} \corr
\left(\begin{smallmatrix}
0&0&0&0\\ 0&0&0&0\\ 0&0&1&0\\ 0&0&0&0
\end{smallmatrix}\right).
\end{align}
These density matrices are orthogonal, $\tr(\zzu\zxv) = 0$,
because his measurement means allow him to distinguish the two
corresponding preparations in one shot. This is precisely what
Tatiana could not do, instead, and so she represented the two
preparations by the non-orthogonal density matrices $\zz =
\ketbra{\zzp}{\zzp}$ and $\zx = \ketbra{\zzm}{\zzm}$.

Willard mixed the two separable gases with his semi-permeable
\diaphragm s, obtaining work (Fig.~\ref{fig:willard},~b), so that
the container eventually contained a $\zmi$-gas, where
\begin{equation}
\label{eq:mixwill}
\zmi = \frac{1}{2} \zzu 
+ \frac{1}{2} \zxv 
\corr \tfrac{1}{2}\left(\begin{smallmatrix}
1&0&0&0\\ 0&0&0&0\\ 0&0&1&0\\ 0&0&0&0
\end{smallmatrix}\right).
\end{equation}

Tatiana's subsequent separation by means of her semi-permeable
\diaphragm s (Fig.~\ref{fig:willard},~c~d) which can separate the
preparations corresponding to $\za$ and $\zb$,
is represented by Willard by the two-element \POVM\ $\set{\zE^+,
\zE^-}$, with
\begin{gather}
\begin{split}
\zwEF &\defin \ketbra{\zwbBk}{\zwbBk} +
\ketbra{\zwcCk}{\zwcCk},\\
 &\corr \tfrac{1}{4}\left(\begin{smallmatrix}
2\pm\sqrt{2}&\pm\sqrt{2}&0&0\\
\pm\sqrt{2}& 2\mp \sqrt{2}&0&0 \\
0&0&2\pm\sqrt{2}&\pm\sqrt{2}\\
0&0&\pm\sqrt{2}& 2\mp \sqrt{2} \\
\end{smallmatrix}\right),
\end{split}
\\
\intertext{where}
\ket{\zwbB} \defin \Bigl(2\pm
\sqrt{2}\Bigr)^{-\frac{1}{2}} \bigl(\ket{\zzupm} \pm
\ket{\zxupm}\bigr),
\\
\ket{\zwcC} \defin \Bigl(2\pm
\sqrt{2}\Bigr)^{-\frac{1}{2}} \bigl(\ket{\zzvpm} \pm
\ket{\zxvpm}\bigr)
\end{gather}
(\cf\ \eqn~\eqref{eq:newketalf}), and the associated \CPM s
(projections) $\zmi \mapsto \zwEF\zmi\zwEF/ \tr(\zwEF\zmi\zwEF)$.

That separation led to a chamber, with volume $0.854\, V$,
containing the gas \emph{mixture}
\begin{equation}
\label{eq:mixaz}
\frac{1}{2} \ketbra{\zaup}{\zaup}+
\frac{1}{2} \ketbra{\zavp}{\zavp},
\end{equation}
and another chamber, with volume $0.146\, V$, containing the gas
\emph{mixture}
\begin{equation}
\label{eq:mixbz}
\frac{1}{2} \ketbra{\zbup}{\zbup}+
\frac{1}{2} \ketbra{\zbvp}{\zbvp}.
\end{equation}
Note that Tatiana could not notice that these were mixtures,
because of her limited instrumentation. Moreover, from Willard's
point of view the separation brought about a transformation of
the gases' quantum degrees of freedom.

The following step corresponded to the unitary rotations
\begin{align}
\ketbra{\zwbBpk}{\zwbBpk} &\mapsto \ketbra{\zzup}{\zzup},\\
\ketbra{\zwcCpk}{\zwcCpk} &\mapsto
\ketbra{\zzvp}{\zzvp}\\
\intertext{for gases in the upper chamber, and}
\ketbra{\zwbBmk}{\zwbBmk} &\mapsto \ketbra{\zzup}{\zzup},\\
\ketbra{\zwcCmk}{\zwcCmk} &\mapsto
\ketbra{\zzvp}{\zzvp}
\end{align}
for the gases in the lower chamber. The successive elimination
and reinsertion of the impermeable \diaphragm\ yielded two
chambers of equal volume $V/2$ and same content, \viz\ the
\emph{mixture} of gases described by $\ketbra{\zzup}{\zzup}$ and
$\ketbra{\zzvp}{\zzvp}$ (Fig.~\ref{fig:willard},~e).

Tatiana's final rotation for the gas in the lower chamber,
\begin{align}
\ket{\zzup} &\mapsto \ket{\zxup},&
\ket{\zzvp} &\mapsto \ket{\zxvp},
\end{align}
only led to two equal chambers containing the \emph{mixtures} of
$\tfrac{1}{2} \ketbra{\zzup}{\zzup}+ \tfrac{1}{2}
\ketbra{\zzvp}{\zzvp}$ and $\tfrac{1}{2} \ketbra{\zxup}{\zxup}+
\tfrac{1}{2} \ketbra{\zxvp}{\zxvp}$ gases respectively
(Fig.~\ref{fig:willard},~f).

From her point of view, \ie, from what her measuring means could
tell, this final situation was identical with the initial one
(Fig.~\ref{fig:willard},~a), \ie\ a $\zz$-gas in one chamber and
an $\zx$-gas in the other, and so she thought the thermodynamic
cycle completed. But we see now that the final and initial
situations were in fact \emph{different}. Hence the second law
was \emph{not} violated, because \emph{the cycle was not
completed}, and the form~\eqref{eq:seclawcycl} for the second law
\emph{does not apply}.

It is also easy to see that in order to return to the initial
condition an amount of work $-Q''' \ge 4\times (1/4) \zN \zk \zT
\ln 2 \approx 0.693\, \zN \zk \zT$ has to be \emph{spent} to
separate the $\ketbra{\zzup}{\zzup}$-gas from the
$\ketbra{\zzvp}{\zzvp}$-gas, and analogously for the
$\ketbra{\zxup}{\zxup}$- and $\ketbra{\zxvp}{\zxvp}$-gases. A
final operation must then be performed corresponding to the
rotations of the density matrices $\ketbra{\zzvp}{\zzvp}$ and
$\ketbra{\zxup}{\zxup}$ to $\ketbra{\zzup}{\zzup}$ and
$\ketbra{\zxvp}{\zxvp}$ respectively, and we have finally reached
again the initial condition (Fig.~\ref{fig:willard},~a). The total
amount of heat \emph{absorbed} by the gases in this completing
process would then be
\begin{multline}\label{eq:totalabsperes}
Q =Q' + Q'' + Q''' \le
(0.693 - 0.416 - 0.693) \zN
\zk \zT = {}\\
-0.416 \zN \zk \zT \le 0,
\end{multline}
and the second law, for the completed cycle, would be satisfied
(strictly so: we see that the whole process is irreversible, and
it is easy to check that the only irreversible step was Tatiana's
transformation and separation into $\za$- and $\zb$-gases).

The original conclusion has thus been reversed: no violation of
the second law is found.

\section{Conclusions}
\label{sec:concldisc}

Preparation procedure and density matrix, with their respective
properties, are quite different concepts, but intimately related
because the latter is a mathematical representation of the
former. In particular, the orthogonality of two density matrices
\emph{mathematically represents} the fact that their
corresponding preparation procedures can be distinguished in one
shot. The one-shot distinguishability of two preparation
procedures is also equivalent to the possibility of separating
them by appropriate semi-permeable \diaphragm s. This equivalence
is important because it relates their statistical and gross
thermodynamic properties.

A consequence of this physical-mathematical relation is that an
observer cannot assume the non-orthogonality of two density
matrices and, at the same time, claim the one-shot
distinguishability of their corresponding preparation procedures,
because this assumption is self-contradictory and can only lead
to vain
conclusions. We have seen this kind of contradictory assumption
behind the second statement~\eqref{eq:second}.

However, \emph{two} observers can represent the same preparations
by means of density matrices having different numerical values
and properties (\eg, orthogonality), and even different
dimensionality. This may happen either because the two observers
have different pieces of knowledge about the preparations'
properties and the measurements available; or because each
observer, having a different purpose or application, considers
different sets of preparations and measurements to describe the
same physical phenomenon. In particular, properties like one-shot
distinguishability --- and thus also orthogonality --- always
depend on the particular set of measurement procedures which an
observer decides to consider (sometimes the whole known set at
his or her disposal).

We have seen an example of this point in Peres'
thought-experiment: An observer does not know of any measurement
procedure that can distinguish in one shot two particular
preparations; accordingly, she represents them by non-orthogonal
density matrices. A second observer then shows the existence of
such a measurement procedure, and so both observers \emph{must}
then use orthogonal density matrices to represent the
preparations, at least when they want to operate with that
measurement.

In any case, the consistency between the physical phenomena
considered and their mathematical description is always
essential.

\begin{acknowledgements}
We thank an anonymous reviewer (see Footnote~\ref{fn:reviewer2})
for some constructive criticisms which helped to explain the
ideas of this paper more clearly. PGLM thanks Louise for
encouragement, suggestions, and invaluable odradek, and Peter
Morgan for many useful comments. Special and affectionate thanks
also to the staff of the Royal Institute of Technology's Library,
for their irreplaceable work and ever prompt support. AM thanks 
Anders Karlsson for encouragement.

We dedicate this paper to the memory of Asher Peres and to his
clarifying work in quantum mechanics.
\end{acknowledgements}

\appendix*\section{Orthogonality proof}

Suppose that two preparation procedures are represented by
density matrices $\zphi$ and $\zpsi$, and that we have a
measurement procedure represented by a \POVM\ $\set{\zE_\mu,
\zF_\nu}$ such that
\begin{equation}
\begin{split}
&\tr(\zphi\zE_\mu) = 0\quad\text{and}\quad\tr(\zpsi\zE_\mu) \neq 0
\qquad\text{for all $\zE_\mu$},\\
&\tr(\zpsi\zF_\nu) = 0\quad\text{and}\quad\tr(\zphi\zF_\nu) \neq 0
\qquad\text{for all $\zF_\nu$}.
\end{split}\label{eq:disting-bis}
\end{equation}
These equations mathematically represent the condition of
one-shot distinguishability, as discussed in
\sect~\ref{sec:distorth}.

We want to prove that from the above equations it follows that
the two density matrices are orthogonal:
\begin{equation}
\tr(\zphi \zpsi) = 0.\label{eq:thesisortho}
\end{equation}

To prove this, let us consider the two-element \POVM\ $\set{\zE,
\zF}$ given by:
\begin{align}\label{eq:coarsegr}
\zE &\defin \sum_\mu \zE_\mu,
&\zF &\defin \sum_\nu \zF_\nu.
\end{align}
This \POVM\ represents a ``coarse graining'' of the original
measurement procedure, with some results grouped together. It is
easy to see that it satisfies
\begin{equation}
\begin{split}
&\tr(\zphi\zE) = 0\quad\text{and}\quad\tr(\zpsi\zE) = 1 \neq 0,\\
&\tr(\zpsi\zF) = 0\quad\text{and}\quad\tr(\zphi\zF) = 1 \neq 0.
\end{split}\label{eq:disting-two-bis}
\end{equation}

The first \POVM\ element $\zE$ can be written as a sum of
one-dimensional orthogonal projectors $\set{\proj{i}}$ (here and
in the following the indices $i$, $k$, $j$, and $l$ take values
in appropriate and possibly distinct ranges):
\begin{equation}
\label{eq:decE}
\zE = \sum_{i} e_i \proj{i},
\quad\text{with $0 < e_i \le 1$ for all $i$} 
\end{equation}
(note that the sum is not necessarily a convex combination, even
though its coefficients are positive and not greater than
unity).

We can decompose the first density matrix
$\zphi$ in an analogous manner:
\begin{equation}
\label{eq:decD1}
\zphi = \sum_{k} \phi_{k} \proj{\Tilde{k}},
\quad\text{with $0 < \phi_{k} \le 1$ for all $k$
 and $\sum_{k} \phi_{k} =1$},
\end{equation}
where the projectors $\set{\proj{\Tilde{k}}}$ are not
necessarily parallel or orthogonal to the
$\set{\proj{i}}$.

The assumption $\tr(\zphi\zE) = 0$ can be written as
\begin{equation}
\label{eq:assorth1}
\tr(\zphi\zE) =
\sum_{k,i}
\phi_{k} e_i
\abs{\braket{\Tilde{k}}{i}}^2 = 0,
\end{equation}
which by the strict positivity of the $\phi_{k}$ and
$e_i$ implies
\begin{equation}
\label{eq:assorth1b}
\braket{\Tilde{k}}{i} = 0,
\quad\text{for all $k$ and $i$,}
\end{equation}
\ie, the projectors --- equivalently, the eigenvectors ---
of the matrix $\zE$ are in fact all orthogonal to those of
the matrix $\zphi$.

This means that the $\set{\proj{i}}$ can be completed by the
$\set{\proj{\Tilde{k}}}$ and some additional projectors
$\set{\proj{j'}}$ to form a complete set of orthonormal
projectors --- an orthonormal basis:
\begin{equation}
\set{\proj{i},
\proj{\Tilde{k}},
\proj{j'}}_{ikj}.\label{eq:complset}
\end{equation}

Writing the identity matrix $\id$ in terms of the new
projectors,
\begin{equation}
\label{eq:idproj}
\id \equiv 
\sum_{i} \proj{i}
+ \sum_{k} \proj{\Tilde{k}},
+ \sum_{j} \proj{j'},
\end{equation}
we have that the second \POVM\ element $\zF$ must be given
by
\begin{equation}
\label{eq:zF}
\zF = \id - \zE = 
\sum_{i} (1-e_i) \proj{i}
+ \sum_{k} \proj{\Tilde{k}}
+ \sum_{j} \proj{j'}.
\end{equation}

Let us now decompose the second density matrix $\zpsi$ as
\begin{equation}
\label{eq:decD2}
\zpsi = \sum_{l} \psi_{l} \proj{{\Hat{l}}},
\quad\text{with $0 < \psi_{l} \le 1$ for all $l$
 and  $\sum_{l} \psi_{l} =1$},
\end{equation}
where the projectors $\set{\proj{\Hat{l}}}$ do not
necessarily belong to the complete set previously
introduced.

Using \eqns~\eqref{eq:decD2} and~\eqref{eq:zF}, we rewrite
the assumption that $\tr(\zpsi\zF) = 0$ as
\begin{equation}
\label{eq:assorth2}
\begin{split}
\tr(\zpsi\zF) &=
 \sum_{l,i} \psi_{l} (1-e_i)
\abs{\braket{\Hat{l}}{i}}^2 +{}\\
&\quad  \sum_{l,k}
\psi_{l} 
\abs{\braket{\Hat{l}}{\Tilde{k}}}^2 
+
 \sum_{l,j} 
\psi_{l}
\abs{\braket{\Hat{l}}{j'}}^2 = 0.
\end{split}
\end{equation}
Noting that all the sum terms are non-negative, and that
the coefficients $\psi_{l}$ are strictly positive, we
deduce in particular that
\begin{equation}
\label{eq:assorth1c}
\braket{\Hat{l}}{\Tilde{k}} = 0,
\quad \text{ for all $l$ and $k$,}
\end{equation}
which means that the eigenvectors of the density matrix
$\zphi$ are all orthogonal to those of the matrix $\zpsi$.

We have thus
\begin{equation}
\label{eq:tobeprovorth}
\tr(\zphi\zpsi) =
\sum_{k,l}
\phi_{k}\psi_{l}
\abs{\braket{\Hat{l}}{\Tilde{k}}}^2 
= 0,
\end{equation}
\hspace{\stretch{1}}\QED

The converse is also easy to demonstrate. Consider again the
definitions~\eqref{eq:decD1} and~\eqref{eq:decD2}, and assume
that these density matrices are orthogonal, \ie,
\eqn~\eqref{eq:tobeprovorth}. It follows that their eigenvectors
are orthogonal, \eqn~\eqref{eq:assorth1c}, and may be used
together with additional projectors $\set{\proj{j'}}$ to form
a complete orthonormal set
\begin{equation}
\set{\proj{\Tilde{k}},
\proj{\Hat{l}},
\proj{j'}}_{k,l,j}.\label{eq:complset2}
\end{equation}

Now define the operators
\begin{align}
\zE &\defin \sum_{k} \proj{\Tilde{k}},\\
\zF &\defin  \id - \zE = 
+ \sum_{l} \proj{\Hat{l}}
+ \sum_{j} \proj{j'}
\end{align}
(they are \emph{not} those used in the previous proof, although
we use the same symbols for convenience). These operators form,
as is easily checked, a \POVM\ $\set{\zE, \zF}$, which satisfy
\eqns~\eqref{eq:disting-two-bis}. If we moreover assume that any
\POVM\ may be physically realised by some measurement
procedure,\footnote{See, however, Footnote~\ref{fn:existassump}.}
then we have proven the converse statement: if two preparations
are represented by orthogonal density matrices, it means that
they are one-shot distinguishable.\hspace{\stretch{1}}\QED


\providecommand{\href}[2]{#2}
\providecommand{\eprint}[2]{\texttt{\href{#1#2}{#2}}}
\renewcommand{\eprint}[2]{\texttt{\href{#1#2}{#2}}}
\newcommand{\arxiveprint}[1]{arxiv eprint
\eprint{http://arxiv.org/abs/}{#1}}
\newcommand{\mparceprint}[1]{mp\_arc eprint
\eprint{http://www.ma.utexas.edu/mp_arc-bin/mpa?yn=}{#1}}
\newcommand{\philscieprint}[1]{philsci eprint
\eprint{http://philsci-archive.pitt.edu/archive/}{#1}}
\newcommand{\citein}[1]{\textnormal{\citet{#1}}}


(Note: `arxiv eprints' are located at
\url{http://arxiv.org/}.)
\end{document}